\documentclass[twocolumn,tighten]{aastex62}
\hypersetup{linkcolor=red,citecolor=blue,filecolor=cyan,urlcolor=magenta}
\usepackage{amsmath}
\usepackage{natbib}
\usepackage{color}

\newcommand{\degree}{$^{\circ}$}

\newcommand{\kmsec}{\mbox{km~s$^{\rm -1}$}}
\newcommand{\logg}{\mbox{log~{\it g}}}
\newcommand{\msun}{\mbox{$M_{\odot}$}}
\newcommand{\teff}{\mbox{$T_{\rm eff}$}}

\newcommand{\rpro}{\mbox{{\it r}-process}}
\newcommand{\spro}{\mbox{{\it s}-process}}

\newcommand{\rone}{\mbox{{\it r}-I}}
\newcommand{\rtwo}{\mbox{{\it r}-II}}
\newcommand{\rettwolong}{\object[NAME RETICULUM II]{Reticulum~II}}
\newcommand{\rettwo}{\object[NAME RETICULUM II]{Ret~II}}

\def\vector#1{\mbox{\boldmath $#1$}}

\newcommand{\rapo}{r_\ensuremath{\mathrm{apo}}}
\newcommand{\rperi}{r_\ensuremath{\mathrm{peri}}}
\newcommand{\zmax}{Z_\ensuremath{\mathrm{max}}}

\newcommand{\agama}{\texttt{Agama}}

\newcommand{\eq}[1]{\begin{align}#1\end{align}}

\shorttitle{Kinematics of Field $r$-process Enhanced Stars}
\shortauthors{Roederer et al.}
\accepted{for publication in the Astronomical Journal}

\begin{document}

\title{%
Kinematics of Highly {\it R}-Process-Enhanced Field Stars:\ 
Evidence for an Accretion Origin \\
and Detection of Several Groups from Disrupted Satellites
}

\author{Ian U.\ Roederer}
\affiliation{%
Department of Astronomy, University of Michigan,
1085 S.\ University Ave., Ann Arbor, MI 48109, USA}
\affiliation{%
Joint Institute for Nuclear Astrophysics -- Center for the
Evolution of the Elements (JINA-CEE), USA}
\email{Email:\ iur@umich.edu}

\author{Kohei Hattori}
\affiliation{%
Department of Astronomy, University of Michigan,
1085 S.\ University Ave., Ann Arbor, MI 48109, USA}

\author{Monica Valluri}
\affiliation{%
Department of Astronomy, University of Michigan,
1085 S.\ University Ave., Ann Arbor, MI 48109, USA}

\begin{abstract}

We present kinematics of 
35 highly \rpro-enhanced ([Eu/Fe] $\geq +$0.7) 
metal-poor ($-$3.8 $<$ [Fe/H] $< -$1.4)
field stars.
We calculate six-dimensional positions and velocities, 
evaluate energies and integrals of motion, and
compute orbits for each of these stars
using parallaxes and proper motions from the
second \textit{Gaia} data release
and published radial velocities.
All of these stars have halo kinematics.
Most stars (66\%) remain in the inner regions of the halo ($<$~13~kpc),
and many (51\%) have orbits that 
pass within 2.6~kpc of the Galactic center.
Several stars (20\%) have orbits that extend beyond 20~kpc,
including one with an orbital apocenter larger than the
Milky Way virial radius.
We apply three clustering methods to 
search for structure in phase space,
and we identify eight groups.
No abundances are considered in the clustering process,
but the [Fe/H] dispersions of the groups are 
smaller than would be expected by random chance.
The orbital properties, clustering in phase space and metallicity,
and lack of highly \rpro-enhanced stars on disk-like orbits
indicate that such stars likely 
were accreted from disrupted satellites.
Comparison with the galaxy luminosity-metallicity relation
suggests $M_{V} \gtrsim -9$ for most of the progenitor satellites,
characteristic of ultra-faint or low-luminosity classical
dwarf spheroidal galaxies.
Environments with low rates of star formation and Fe production,
rather than the nature of the \rpro\ site,
may be key to obtaining the [Eu/Fe] ratios found in 
highly \rpro-enhanced stars.

\end{abstract}


\section{Introduction}
\label{intro}

The heaviest elements found in many metal-poor stars were
produced by the rapid neutron-capture process (\rpro)
in earlier generations of stars.
Work by \citet{gilroy88} demonstrated that 
genuine differences exist in the overall levels of enhancement
of \rpro\ elements relative to Fe in metal-poor stars.
The recognition of the highly \rpro-enhanced star 
\object[BPS CS 22892-052]{CS~22892-052}
by \citet{sneden94} 
in the HK Survey of \citet{beers92}
erased any lingering doubt about 
the inhomogeneous distribution of \rpro\ elements
in the environments where metal-poor stars formed.
Stars that exhibit Eu/Fe ratios at least 10 times higher than in the Sun,
like \object[BPS CS 22892-052]{CS~22892-052},
comprise only a small fraction 
($\approx$~3\%; \citealt{barklem05heres})
of all metal-poor field stars, and
none are known to be physically associated 
with each other \citep{roederer09a}.

The environmental impact of the \rpro---expressed 
through the occurrence frequency, distribution, and enhancement levels
of \rpro\ elements in stars---can help
associate \rpro\ abundance patterns with their nucleosynthetic origins.
Observations of the kilonova associated with
gravitational wave event GW170817 
\citep{abbott17prl,abbott17multimessenger}
provide the most direct confirmation that neutron-star mergers
are a site capable of producing heavy elements by \rpro\ nucleosynthesis
(e.g., \citealt{cowperthwaite17,drout17,kasen17,tanvir17}).
The occurrence frequency and level of \rpro\ enhancement
of stars in 
dwarf galaxies supports this conclusion
(e.g., \citealt{ji16nat,safarzadeh17ret2,tsujimoto17}),
although those results alone cannot exclude 
rare classes of supernovae as an additional site
(e.g., \citealt{tsujimoto15mrsn,beniamini16b}).
Chemical evolution models \citep{cote18rpro} and
simulations \citep{naiman18}
can help generalize this result
to \rpro\ production in the Milky Way.
The $^{244}$Pu abundance in deep-sea sediments,
which can be used to infer the 
content of this \rpro-only isotope in the ISM,
also points to rare \rpro\ events like
neutron-star mergers
\citep{hotokezaka15,wallner15}.

We lack similar, direct
knowledge of the birth environments
of highly \rpro-enhanced stars in the Milky Way halo field.
An increasing number of these stars are now known
(e.g., \citealt{hansen18}, and other ongoing work by the
\textit{R}-Process Alliance).
Their proximity to the Sun permits
detailed abundance inventories to be derived from
optical, ultraviolet, and near-infrared spectra
(e.g., \citealt{sneden98,roederer12d,afsar16}).
Five-parameter astrometric solutions 
(parallax, right ascension, declination, 
proper motion in right ascension, proper motion in declination)
are now available
for many of these stars 
in the second data release of the \textit{Gaia} mission
(DR2; \citealt{lindegren18}).
Line-of-sight velocities
based on high-resolution optical spectroscopy
are also available for these stars.
The full space motion of each star can be reconstructed from
these six parameters once a Galactic potential is adopted.
We use these data to examine the kinematic properties of a large 
sample of highly \rpro-enhanced field stars
for the first time.

We present our sample of 
highly \rpro-enhanced field stars in Section~\ref{sample},
and we present their astrometric and velocity
data in Section~\ref{kinematicdata}.
We describe our calculations of the kinematics
in Section~\ref{orbits}.
We discuss the implications of these calculations
in Section~\ref{discussion},
and we summarize our conclusions in Section~\ref{conclusions}.
Throughout this work,
we adopt the standard nomenclature:\ 
for elements X and Y, 
[X/Y] is the abundance ratio relative to the
Solar ratio, defined as
$\log_{10} (N_{\rm X}/N_{\rm Y}) -
\log_{10} (N_{\rm X}/N_{\rm Y})_{\odot}$.

\section{Sample Selection}
\label{sample}

Many highly \rpro-enhanced stars have been identified
and analyzed individually over the last 25~years.
Our sample includes stars from the literature 
which show at least moderately high levels
of \rpro\ enhancement relative to Fe,
[Eu/Fe]~$\geq +$0.7; i.e., enhanced by a factor of 5 
relative to the Solar ratio.
Europium (Eu, $Z =$~63) 
is commonly used as a proxy for the overall level
of \rpro\ enhancement in a star.
A large fraction 
($\approx$~94--98\%; \citealt{sneden08,bisterzo11})
of the Eu in the Solar system
originated via the \rpro,
despite the fact that both the \rpro\ and the \spro\
(slow neutron-capture process)
contributed roughly equal amounts to the total
mass of elements heavier than the Fe group
in the Solar system.
We also require that the heavy-element abundance pattern 
in each star has been scrutinized
in sufficient detail to determine that the \rpro\ 
was the dominant source of the heavy elements
(e.g., \citealt{sneden96}).
We only include field stars in our sample,
so \rpro-enhanced stars in dwarf galaxies and globular clusters
are not considered.

Table~\ref{abundtab} lists the 83~stars satisfying our criteria,
along with the metallicity ([Fe/H]), 
europium to iron ratio ([Eu/Fe]),
europium to hydrogen ratio ([Eu/H]),
and the literature references for these abundances.
We only calculate kinematic and orbital properties
for the subset of stars in Table~\ref{abundtab}
with relatively small uncertainties in their
parallax measurements, as discussed in 
Section~\ref{kinematicdata}.
We retain all 83~stars in Table~\ref{abundtab}
as a reference, however, anticipating that better distance estimates
will be available in the future.

We emphasize that the sample in Table~\ref{abundtab}
is subject to strong observational biases.
Most of these stars were recognized
as being \rpro\ enhanced during
high-resolution spectroscopic followup of
metal-poor candidates identified
via objective-prism surveys
\citep{bidelman73,bond80,beers85,beers92,christlieb08}.
Many observers contributed to these
efforts over the decades, and 
their decisions of which stars to observe
are somewhat subjective and not easily quantified.

\startlongtable
\begin{deluxetable}{lcccc}
\tablecaption{List of Known Highly \rpro-Enhanced Field Stars, 
Sorted by Decreasing [Eu/Fe] Ratios
\label{abundtab}}
\tablewidth{0pt}
\tabletypesize{\tiny}
\tablehead{
\colhead{Star} &
\colhead{[Fe/H]} &
\colhead{[Eu/Fe]} &
\colhead{[Eu/H]} &
\colhead{Reference} 
}
\startdata
\object[SDSS J235718.91-005247.8]{J235718.91$-$005247.8}       & $-$3.36 &  1.92 &    $-$1.44 &  \citet{aoki10} \\
\object[HE 1523-0901]{HE~1523$-$0901}                          & $-$2.95 &  1.81 &    $-$1.14 &  \citet{frebel07he} \\
\object[SMSS J175046.30-425506.9]{SMSS~J1750460.30$-$425506.9} & $-$2.17 &  1.75 &    $-$0.42 &  \citet{jacobson15smss} \\
\object[BPS CS 29497-004]{CS~29497-004}                        & $-$2.85 &  1.67 &    $-$1.18 &  \citet{hill17} \\
\object[RAVE J203843.2-002333]{J203843.2$-$002333}             & $-$2.91 &  1.64 &    $-$1.27 &  \citet{placco17rpro} \\
\object[BPS CS 22892-052]{CS~22892-052}                        & $-$3.10 &  1.64 &    $-$1.46 &  \citet{sneden03a} \\
\object[BPS CS 31082-001]{CS~31082-001}                        & $-$2.90 &  1.63 &    $-$1.27 &  \citet{hill02} \\
\object[2MASS J14325334-4125494]{J14325334$-$4125494}          & $-$2.79 &  1.61 &    $-$1.18 &  \citet{hansen18} \\
\object[HE 1226-1149]{HE~1226$-$1149}                          & $-$2.91 &  1.55 &    $-$1.36 &  \citet{cohen13} \\
\object[2MASS J02462013-1518419]{J02462013$-$1518419}          & $-$2.71 &  1.45 &    $-$1.26 &  \citet{hansen18} \\
\object[HE 1219-0312]{HE~1219$-$0312}                          & $-$2.92 &  1.38 &    $-$1.54 &  \citet{hayek09} \\
\object[HD 222925]{HD~222925}                                  & $-$1.47 &  1.33 &    $-$0.14 &  \citet{roederer18d} \\
\object[2MASS J20093393-3410273]{J20093393$-$3410273}          & $-$1.99 &  1.32 &    $-$0.67 &  \citet{hansen18} \\
\object[2MASS J21064294-6828266]{J21064294$-$6828266}          & $-$2.76 &  1.32 &    $-$1.44 &  \citet{hansen18} \\
\object[2MASS J09544277+5246414]{J09544277$+$5246414}          & $-$2.99 &  1.28 &    $-$1.71 &  \citet{holmbeck18rpro} \\
\object[2MASS J15383085-1804242]{J15383085$-$1804242}          & $-$2.09 &  1.27 &    $-$0.82 &  \citet{sakari18a} \\
\object[SMSS J183128.71-341018.4]{SMSS~J183128.71$-$341018.4}  & $-$1.83 &  1.25 &    $-$0.58 &  \citet{howes16} \\
\object[2MASS J21091825-1310062]{J21091825$-$1310062}          & $-$2.40 &  1.25 &    $-$1.15 &  \citet{hansen18} \\
\object[HE 0432-0923]{HE~0432$-$0923}                          & $-$3.19 &  1.25 &    $-$1.94 &  \citet{barklem05heres} \\
\object[BPS CS 31078-018]{CS~31078-018}                        & $-$2.84 &  1.23 &    $-$1.61 &  \citet{lai08} \\
\object[HE 0430-4901]{HE~0430$-$4901}                          & $-$2.72 &  1.16 &    $-$1.56 &  \citet{barklem05heres} \\
\object[BPS CS 22183-031]{CS~22183-031}                        & $-$2.93 &  1.16 &    $-$1.77 &  \citet{honda04b} \\
\object[2MASS J23362202-5607498]{J23362202$-$5607498}          & $-$2.06 &  1.14 &    $-$0.92 &  \citet{hansen18} \\
\object[SMSS J155430.57-263904.8]{SMSS~J155430.57$-$263904.8}  & $-$2.61 &  1.14 &    $-$1.47 &  \citet{jacobson15smss} \\
\object[BPS CS 22945-058]{CS~22945-058}                        & $-$2.71 &  1.13 &    $-$1.58 &  \citet{roederer14e} \\
\object[BPS CS 22945-017]{CS~22945-017}                        & $-$2.73 &  1.13 &    $-$1.60 &  \citet{roederer14e} \\
\object[2MASS J02165716-7547064]{J02165716$-$7547064}          & $-$2.50 &  1.12 &    $-$1.38 &  \citet{hansen18} \\
\object[SMSS J183225.29-334938.4]{SMSS~J183225.29$-$334938.4}  & $-$1.74 &  1.08 &    $-$0.66 &  \citet{howes16} \\
\object[2MASS J19161821-5544454]{J19161821$-$5544454}          & $-$2.35 &  1.08 &    $-$1.27 &  \citet{hansen18} \\
\object[HE 1127-1143]{HE~1127-1143}                            & $-$2.73 &  1.08 &    $-$1.65 &  \citet{barklem05heres} \\
\object[2MASS J17225742-7123000]{J17225742$-$7123000}          & $-$2.42 &  1.07 &    $-$1.35 &  \citet{hansen18} \\
\object[SMSS J062609.83-590503.2]{SMSS~J062609.83$-$590503.2}  & $-$2.77 &  1.06 &    $-$1.71 &  \citet{jacobson15smss} \\
\object[2MASS J18024226-4404426]{J18024226$-$4404426}          & $-$1.55 &  1.05 &    $-$0.50 &  \citet{hansen18} \\
\object[HE 2224+0143]{HE~2224$+$0143}                          & $-$2.58 &  1.05 &    $-$1.53 &  \citet{barklem05heres} \\
\object[BPS CS 22953-003]{CS~22953-003}                        & $-$2.84 &  1.05 &    $-$1.79 &  \citet{francois07} \\
\object[SMSS J175738.37-454823.5]{SMSS~J175738.37$-$454823.5}  & $-$2.46 &  1.02 &    $-$1.44 &  \citet{jacobson15smss} \\
\object[BPS CS 22958-052]{CS~22958-052}                        & $-$2.42 &  1.00 &    $-$1.42 &  \citet{roederer14e} \\
\object[SMSS J024858.41-684306.4]{SMSS~J024858.41$-$684306.4}  & $-$3.71 &  1.00 &    $-$2.71 &  \citet{jacobson15smss} \\
\object[2MASS J18174532-3353235]{J18174532$-$3353235}          & $-$1.67 &  0.99 &    $-$0.68 &  \citet{johnson13} \\
\object[HE 2301-4024]{HE~2301$-$4024}                          & $-$2.11 &  0.98 &    $-$1.13 &  \citet{barklem05heres} \\
\object[HE 2327-5642]{HE~2327$-$5642}                          & $-$2.78 &  0.98 &    $-$1.80 &  \citet{mashonkina10} \\
\object[BPS CS 29491-069]{CS~29491-069}                        & $-$2.55 &  0.96 &    $-$1.59 &  \citet{hayek09} \\
SMSS~J181505.16$-$385514.9                                     & $-$3.29 &  0.96 &    $-$2.33 &  \citet{howes15} \\
\object[HE 2244-1503]{HE~2244$-$1503}                          & $-$2.88 &  0.95 &    $-$1.93 &  \citet{barklem05heres} \\
\object[SMSS J051008.62-372019.8]{SMSS~J051008.62$-$372019.8}  & $-$3.20 &  0.95 &    $-$2.25 &  \citet{jacobson15smss} \\
\object[SMSS J221448.33-453949.9]{SMSS~J221448.33$-$453949.9}  & $-$2.56 &  0.94 &    $-$1.62 &  \citet{jacobson15smss} \\
\object[HE 1044-2509]{HE~1044$-$2509}                          & $-$2.89 &  0.94 &    $-$1.95 &  \citet{barklem05heres} \\
\object[2MASS J19014952-4844359]{J19014952$-$4844359}          & $-$1.87 &  0.93 &    $-$0.94 &  \citet{hansen18} \\ 
\object[BPS CS 22875-029]{CS~22875-029}                        & $-$2.69 &  0.92 &    $-$1.77 &  \citet{roederer14e} \\
\object[BD +173248]{BD $+$17\degree3248}                       & $-$2.06 &  0.91 &    $-$1.15 &  \citet{cowan02} \\
\object[2MASS J19324858-5908019]{J19324858$-$5908019}          & $-$1.93 &  0.90 &    $-$1.03 &  \citet{hansen18} \\
\object[BPS CS 29529-054]{CS~29529-054}                        & $-$2.75 &  0.90 &    $-$1.85 &  \citet{roederer14e} \\
\object[2MASS J21224590-4641030]{J21224590$-$4641030}          & $-$2.96 &  0.90 &    $-$2.06 &  \citet{hansen18} \\
\object[2MASS J15582962-1224344]{J15582962$-$1224344}          & $-$2.54 &  0.89 &    $-$1.65 &  \citet{hansen18} \\
\object[SMSS J063447.15-622355.0]{SMSS~J063447.15$-$622355.0}  & $-$3.41 &  0.89 &    $-$2.52 &  \citet{jacobson15smss} \\
\object[HE 1131+0141]{HE~1131$+$0141}                          & $-$2.48 &  0.87 &    $-$1.61 &  \citet{barklem05heres} \\
\object[2MASS J00405260-5122491]{J00405260$-$5122491}          & $-$2.11 &  0.86 &    $-$1.25 &  \citet{hansen18} \\
\object[BPS CS 22888-047]{CS~22888-047}                        & $-$2.54 &  0.86 &    $-$1.68 &  \citet{roederer14e} \\
\object[BPS CS 22943-132]{CS~22943-132}                        & $-$2.67 &  0.86 &    $-$1.81 &  \citet{roederer14e} \\
\object[BPS CS 22896-154]{CS~22896-154}                        & $-$2.69 &  0.86 &    $-$1.83 &  \citet{francois07} \\
\object[BPS CS 30306-132]{CS~30306-132}                        & $-$2.42 &  0.85 &    $-$1.57 &  \citet{honda04b} \\
\object[BPS CS 22886-012]{CS~22886-012}                        & $-$2.61 &  0.85 &    $-$1.76 &  \citet{roederer14e} \\
\object[HD 115444]{HD~115444}                                  & $-$2.96 &  0.85 &    $-$2.11 &  \citet{westin00} \\
SMSS~J183647.89$-$274333.1                                     & $-$2.48 &  0.82 &    $-$1.66 &  \citet{howes15} \\
\object[BPS CS 22882-001]{CS~22882-001}                        & $-$2.62 &  0.81 &    $-$1.81 &  \citet{roederer14e} \\
\object[HE 2252-4225]{HE~2252$-$4225}                          & $-$2.63 &  0.81 &    $-$1.82 &  \citet{mashonkina14he} \\
\object[HD 20]{HD~20}                                          & $-$1.58 &  0.80 &    $-$0.78 &  \citet{barklem05heres} \\
\object[HD 221170]{HD~221170}                                  & $-$2.18 &  0.80 &    $-$1.38 &  \citet{ivans06} \\
\object[HE 0420+0123a]{HE~0420$+$0123a}                        & $-$3.03 &  0.79 &    $-$2.24 &  \citet{hollek11} \\
\object[HE 0300-0751]{HE~0300$-$0751}                          & $-$2.27 &  0.77 &    $-$1.50 &  \citet{barklem05heres} \\
\object[2MASS J21095804-0945400]{J21095804$-$0945400}          & $-$2.73 &  0.77 &    $-$1.96 &  \citet{hansen18} \\
\object[2MASS J19232518-5833410]{J19232518$-$5833410}          & $-$2.08 &  0.76 &    $-$1.32 &  \citet{hansen18} \\
\object[2MASS J19215077-4452545]{J19215077$-$4452545}          & $-$2.56 &  0.74 &    $-$1.80 &  \citet{hansen18} \\
\object[SMSS J195931.70-643529.3]{SMSS~J195931.70$-$643529.3}  & $-$2.58 &  0.74 &    $-$1.84 &  \citet{jacobson15smss} \\
\object[2MASS J17435113-5359333]{J17435113$-$5359333}          & $-$2.24 &  0.73 &    $-$1.51 &  \citet{hansen18} \\
\object[HE 0240-0807]{HE~0240$-$0807}                          & $-$2.68 &  0.73 &    $-$1.95 &  \citet{barklem05heres} \\
\object[BPS BS 17569-049]{BS~17569-049}                        & $-$2.88 &  0.72 &    $-$2.16 &  \citet{francois07} \\
\object[HE 1430+0053]{HE~1430$+$0053}                          & $-$3.03 &  0.72 &    $-$2.31 &  \citet{barklem05heres} \\
\object[BPS CS 30315-029]{CS~30315-029}                        & $-$3.33 &  0.72 &    $-$2.61 &  \citet{barklem05heres} \\
\object[2MASS J01530024-3417360]{J01530024$-$3417360}          & $-$1.50 &  0.71 &    $-$0.79 &  \citet{hansen18} \\
\object[2MASS J15271353-2336177]{J15271353$-$2336177}          & $-$2.15 &  0.70 &    $-$1.45 &  \citet{hansen18} \\
\object[2MASS J18295183-4503394]{J18294122$-$4504000}          & $-$2.48 &  0.70 &    $-$1.78 &  \citet{hansen18} \\
\object[SMSS J182601.24-332358.3]{SMSS~J182601.24$-$332358.3}  & $-$2.83 &  0.70 &    $-$2.13 &  \citet{howes16} \\
\enddata
\end{deluxetable}

Our definition of a highly \rpro-enhanced star
differs slightly from that found in the literature.
The common ``\rtwo'' designation \citep{beers05}
refers to stars with
[Eu/Fe]~$> +$1.0 and [Ba/Eu]~$<$~0, while the
``\rone'' designation refers to 
stars with more moderate enhancement,
$+$0.3~$\leq$~[Eu/Fe]~$\leq +$1.0 and [Ba/Eu]~$<$~0.
The boundaries defining these classifications are arbitrary.
Our choice of [Eu/Fe]~$\geq +$0.7 is motivated by the
upper envelope of [Eu/Fe]~$\approx +$0.6 found in most
globular clusters (e.g., \citealt{gratton04}), 
kinematically-selected samples of
disk stars (e.g., \citealt{venn04,battistini16}),
and stars toward the Galactic bulge (e.g., \citealt{johnson12}).
[Eu/Fe]~$\geq +$0.7 may represent a more natural and 
physically interpretable 
lower limit for the class of highly \rpro-enhanced stars
than the [Eu/Fe]~$> +$1.0 criterion commonly adopted for \rtwo\ stars,
as we discuss in Section~\ref{environment}.

\begin{figure}
\begin{center}
\includegraphics[angle=0,width=3.35in]{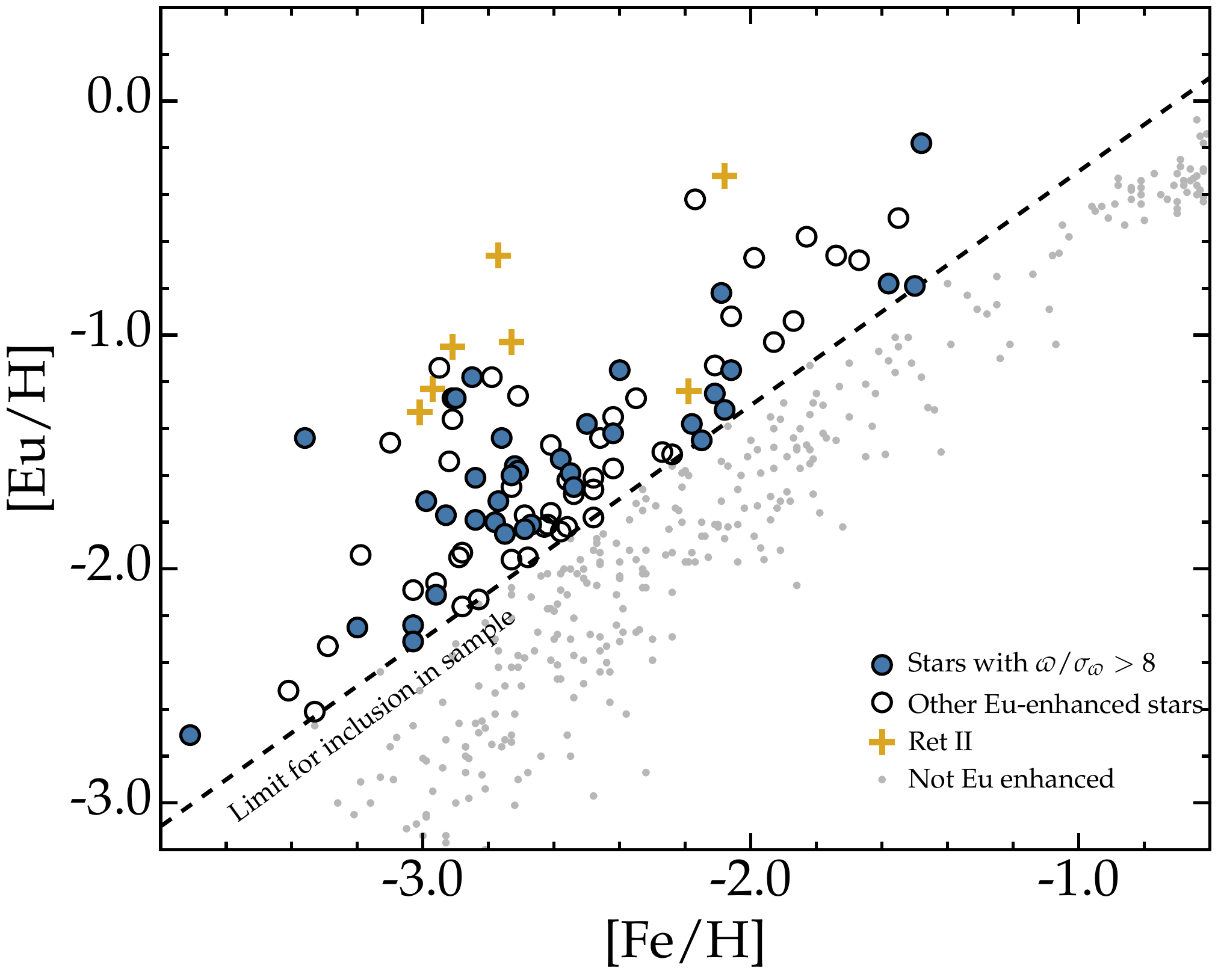} \\
\vspace*{0.1in}
\includegraphics[angle=0,width=3.35in]{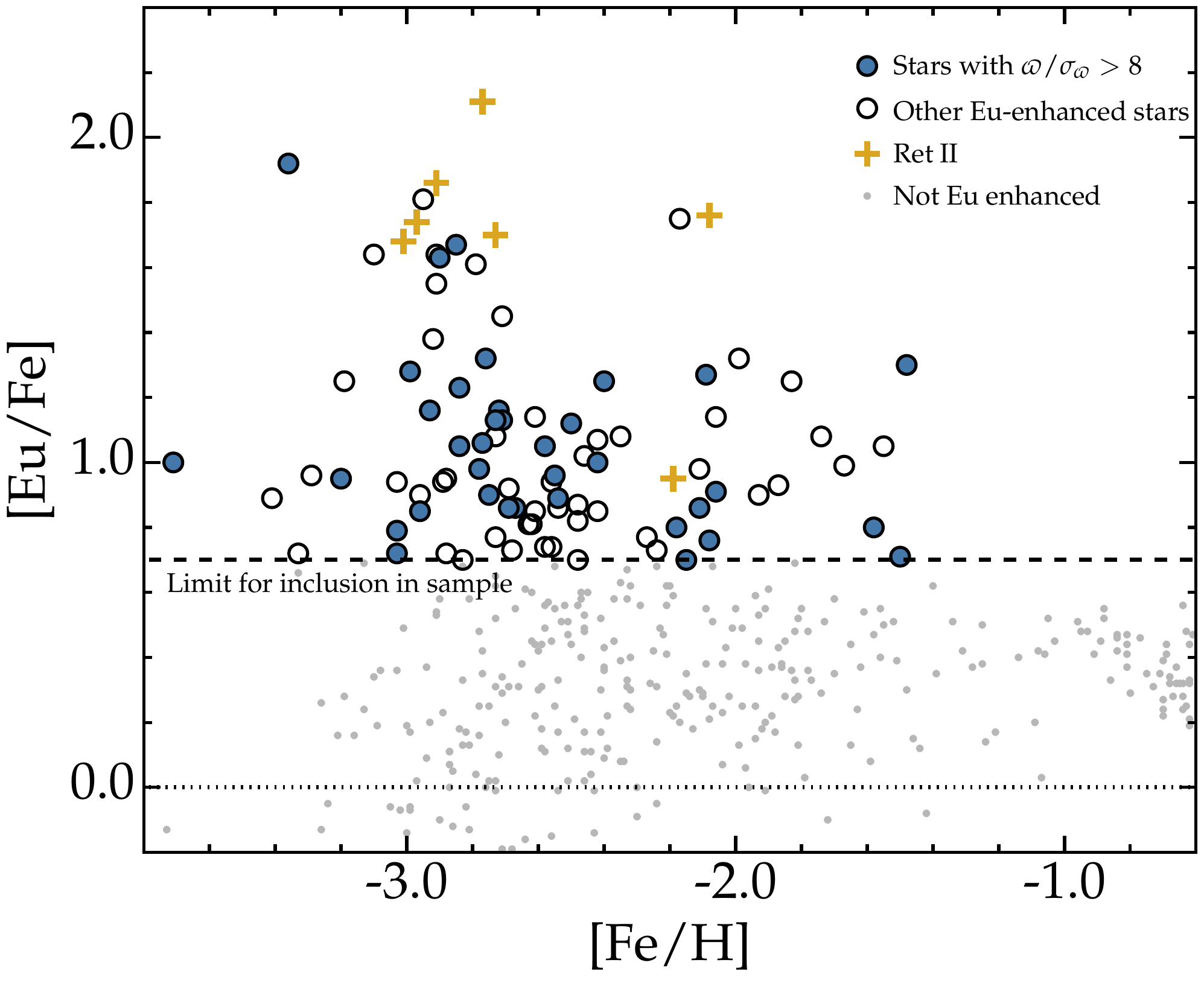}
\end{center}
\caption{
\label{euplot}
The [Eu/H] ratio (top) and [Eu/Fe] ratio (bottom)
as a function of [Fe/H] for stars included in
our sample.
Stars meeting our parallax requirement,
$\varpi/\sigma_{\varpi} \geq$~8.0, 
are indicated by filled circles.
All other highly \rpro-enhanced 
stars are indicated by open circles.
The yellow crosses mark stars in the
\rettwo\ UFD galaxy.
The small gray points mark other stars
from \citet{barklem05heres}, \citet{roederer14c},
\citet{jacobson15smss}, \citet{battistini16}, and
\citet{hansen18}.
The dashed line marks the lower limit of [Eu/Fe]~$\geq +$0.7
for inclusion in our sample.
The dotted line in the bottom panel marks the Solar [Eu/Fe] ratio.
}
\end{figure}

\begin{figure}
\begin{center}
\includegraphics[angle=0,width=2.9in]{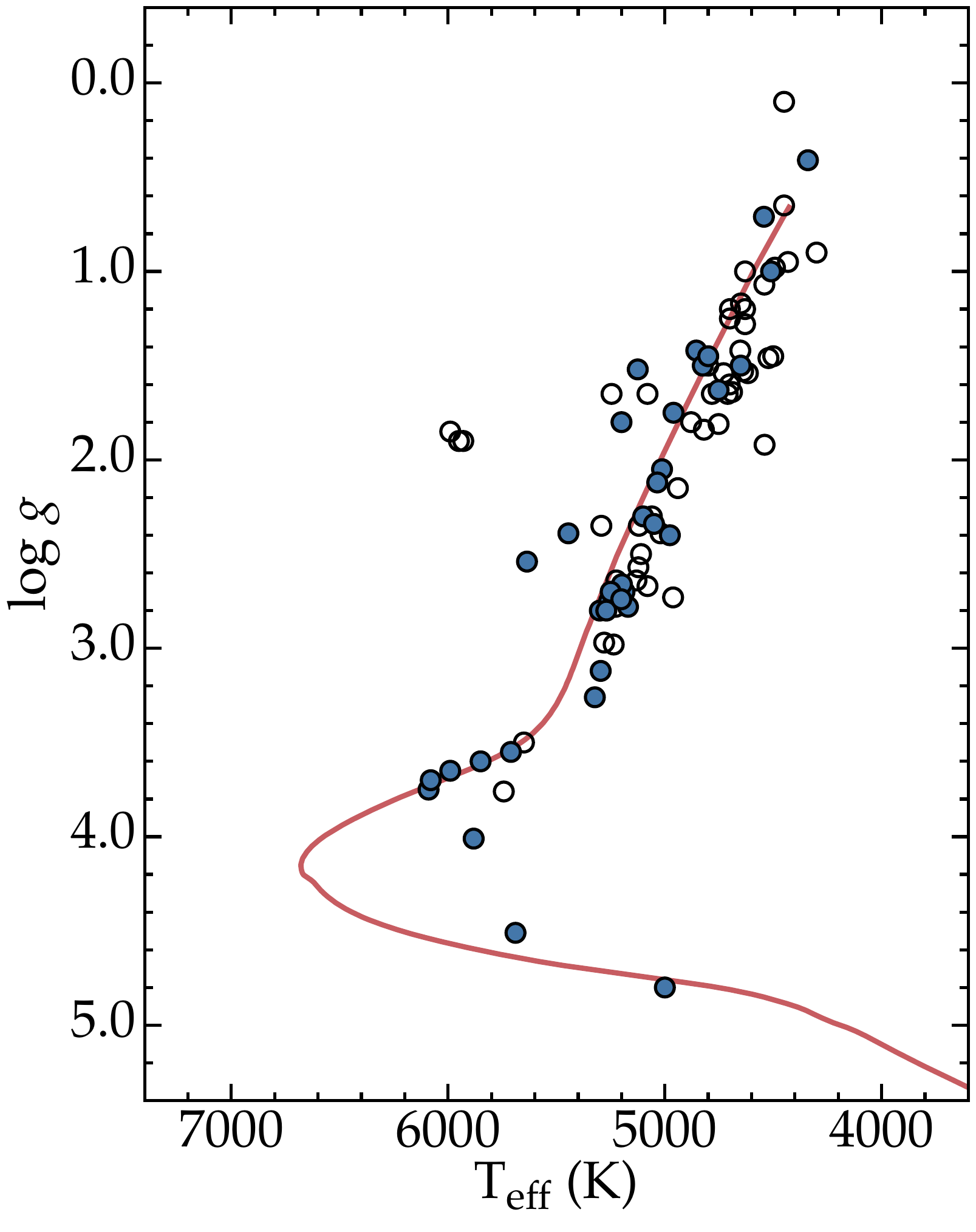} \\
\end{center}
\caption{
\label{cmdplot}
Location of the 83~stars in our sample on a \teff-\logg\ diagram.
Stars meeting our parallax requirement,
$\varpi/\sigma_{\varpi} \geq$~8.0, 
are indicated by filled circles.
All other highly \rpro-enhanced 
stars are indicated by open circles.
The red line is a 13~Gyr isochrone for a 
metal-poor ([Fe/H]~$= -$2.5), 
$\alpha$-enhanced ([$\alpha$/Fe]~$= +$0.4)
stellar population with a standard He mass fraction
(0.2452)
downloaded from the Dartmouth Stellar Evolution Database \citep{dotter08}.
}
\end{figure}

Figure~\ref{euplot} displays the [Eu/H] and [Eu/Fe] ratios
for stars in our sample.
[Eu/H] provides an estimate of the 
amount of dilution of \rpro\ material into gas,
and [Eu/Fe] provides an estimate of the
amount of \rpro\ material relative to the amount of
Fe and other metals from supernovae.
The \rpro-enhanced stars in \rettwolong\
(\rettwo; \citealt{ji16ret2,roederer16b}),
the one known highly \rpro-enhanced galaxy,
are shown for comparison.
These stars
span a range of $\approx$~1~dex in [Eu/H], [Eu/Fe], and [Fe/H],
indicating that stars within an individual dwarf galaxy
exhibit different levels of \rpro\ enrichment
from a single \rpro\ event.
Figure~\ref{cmdplot} illustrates a 
\teff-\logg\ diagram for the stars in our sample.
These stars span a range of evolutionary states,
from main sequence stars to the red horizontal branch.

\section{Input Kinematic Data}
\label{kinematicdata}

Table~\ref{gaiatab} lists the 
\textit{Gaia} DR2 source ID,
parallax ($\varpi$), proper motions ($\mu_{\alpha}\cos\delta$, $\mu_{\delta}$),
distance,
heliocentric radial velocity (RV), and 1$\sigma$ uncertainties for
each of these quantities.
The $\varpi$, $\mu_{\alpha}\cos\delta$, and $\mu_{\delta}$ values
are adopted from \textit{Gaia} DR2
\citep{lindegren18}.
The distances reported in Table~\ref{gaiatab}
are adopted from \citet{bailerjones18} and 
are based on \textit{Gaia} DR2 parallaxes.
Literature references are given for the RV measurements
from high-resolution optical spectroscopy.
We estimate the RV uncertainty based on the
data quality when previous studies did not 
explicitly state this value.
The systemic RV is listed for
known RV-variable stars, when available (e.g., \citealt{hansen15rpro}).
The \textit{Gaia} RV measurements agree with literature values 
to within $\approx$~2--3~\kmsec.

\begin{figure}
\begin{center}
\includegraphics[angle=0,width=3.35in]{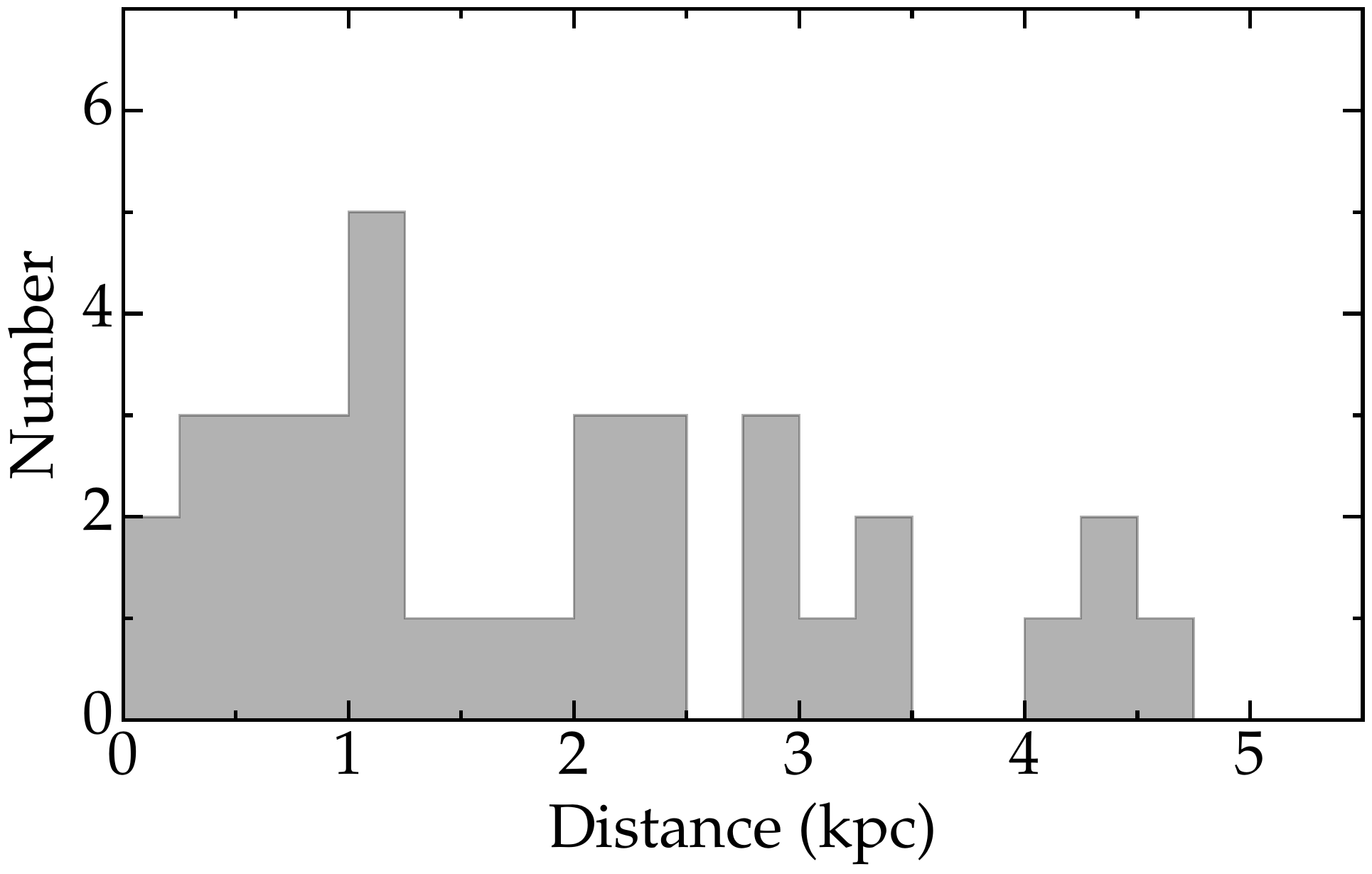}
\end{center}
\caption{
\label{distanceplot}
Histogram of estimated distances \citep{bailerjones18}
to the 35 highly \rpro-enhanced stars in our sample.
}
\end{figure}

We impose a parallax cut on our sample, requiring that
$\varpi/\sigma_{\varpi} \geq$~8.0
(i.e., 12.5\% errors or better).
We determine this value empirically, and
larger parallax uncertainties generally yield
orbital properties
and integrals of motion (Section~\ref{orbits})
that are highly uncertain.
Thirty-five stars pass this cut for further examination.
These stars represent a local sample,
as shown in Figure~\ref{distanceplot}.
The median distance is 1.6~kpc, and 80\% of the sample is
located within 3~kpc of the Sun.

\section{Energy, Actions, and Orbital Calculations}
\label{orbits}

We convert the observed astrometric quantities
into orbital parameters and integrals of motion
for each star in our sample.
We assume that the Sun is on the Galactic plane 
\citep{2017MNRAS.470.1360B}
and is 
$R_0 =$~8.0~kpc away from the Galactic center. 
We also assume that the circular velocity at the Solar position is
$v_0 =$~220~\kmsec\ and the Solar peculiar velocity relative to the 
circular velocity is 
$(U_\odot, V_\odot, W_\odot) =$ (11.1, 12.24, 7.25)~\kmsec\ 
\citep{2010MNRAS.403.1829S}. 
We adopt the realistic gravitational potential model
\texttt{MWPotential2014} \citep{2015ApJS..216...29B}, 
which assumes a virial mass of 
$0.8 \times 10^{12}$~\msun. 
Our calculations account for the correlations between 
$\mu_{\alpha}\cos\delta$ and $\mu_{\delta}$,
$\varpi$ and $\mu_{\alpha}\cos\delta$, and
$\varpi$ and $\mu_{\delta}$
as reported by \textit{Gaia} DR2.

We sample $10^{3}$ sets of 
($\varpi$, $\ell$, $b$, RV, $\mu_{\alpha}\cos\delta$, $\mu_{\delta}$)
from the error distribution of each quantity for each star,
where the uncertainties in $\ell$ and $b$ 
(Galactic longitude and latitude, respectively) are negligible.
This exercise yields $10^{3}$ samples of the six-dimensional 
positions and velocities for each star.
We use the publicly-available \agama\ code
\citep{Vasiliev2018} to
calculate the corresponding stellar orbits
over 3~Gyr for each sample,
yielding 
pericentric radius ($\rperi$), 
apocentric radius ($\rapo$), 
maximum height above or below the Galactic midplane ($\zmax$),
and eccentricity
($e = (\rapo - \rperi) / (\rapo + \rperi)$).
We also use \agama,
which implements an efficient algorithm by \citet{Binney2012},
to evaluate the integrals of motion.
This yields 
the specific orbital energy 
($E = (1/2) \vector{v}^2 + \Phi(\vector{x})$;
hereafter ``energy'')
and three-dimensional action ($\vector{J}=(J_r, J_\phi, J_z)$).
We define the radial and vertical actions, $J_r$ and $J_z$, 
in the same manner as \citet{Binney2012}. 
$J_r$ is defined to be non-negative,
and its value can be interpreted as 
the extent of the radial excursion of an orbit.
For a given $E$, 
$J_r=0$ for circular orbits or shell-like orbits, 
and $J_r$ is large for eccentric orbits. 
$J_z$ is also non-negative, 
and its value can be interpreted as 
the extent of the vertical excursion of an orbit.
For example, $J_z=0$ for planar orbits, 
and $J_z$ is large for orbits with large $\zmax$. 
We define the azimuthal action by
\eq{
J_\phi &= \frac{1}{2 \pi} \oint_{\rm orbit} \mathrm{d}\phi \; R V_\phi = - L_z, 
}
such that prograde stars have $J_\phi > 0$ and $V_\phi > 0$. 
Note that $\pi \simeq 3.14$ is a mathematical constant,
and $\varpi$ denotes the parallax.

Table~\ref{cylindricalvelocitytab} lists, for each star
meeting our $\varpi/\sigma_{\varpi} \geq$~8.0 requirement,
the calculated median
velocities in a cylindrical coordinate system
($V_{R}$, $V_{\phi}$, $V_{z}$) and
$V_{\perp}$, defined as 
$(V_{R}^{2} + V_{z}^{2})^{1/2}$.
Table~\ref{kinematictab} lists
the calculated median
actions ($J_{r}$, $J_{\phi}$, $J_{z}$) and 
energy ($E$).
Table~\ref{orbittab} lists
the calculated median values for
$r_{\rm peri}$, $r_{\rm apo}$, $\zmax$, and $e$.
The columns indicated by a minus or plus sign
represent the difference between the median and the 16$^{\rm th}$
and 84$^{\rm th}$ percentiles 
(analogous to the 1$\sigma$ range) of each quantity.

\begin{deluxetable*}{lccccccccccccccccc}
\rotate
\tablecaption{Parallaxes, Proper Motions, Distances, and Radial Velocities
\label{gaiatab}}
\tablewidth{0pt}
\tabletypesize{\tiny}
\tablehead{
\colhead{Star} &
\colhead{\textit{Gaia} DR2 source ID} &
\colhead{$\varpi$} &
\colhead{Unc.} &
\colhead{} &
\colhead{$\mu_{\alpha} \cos\delta$} &
\colhead{Unc.} &
\colhead{} &
\colhead{$\mu_{\delta}$} &
\colhead{Unc.} &
\colhead{} &
\colhead{Distance} &
\colhead{$-$} &
\colhead{$+$} &
\colhead{} &
\colhead{RV} &
\colhead{Unc.} &
\colhead{RV Reference} \\
\cline{3-4}
\cline{6-7}
\cline{9-10}
\cline{12-14}
\cline{16-17}
\colhead{} &
\colhead{} &
\multicolumn{2}{c}{(mas)} &
\colhead{} &
\multicolumn{2}{c}{(mas~yr$^{-1}$)} &
\colhead{} &
\multicolumn{2}{c}{(mas~yr$^{-1}$)} &
\colhead{} &
\multicolumn{3}{c}{(kpc)} &
\colhead{} &
\multicolumn{2}{c}{(\kmsec)} &
\colhead{} 
}
\startdata
\object[SDSS J235718.91-005247.8]{J235718.91$-$005247.8}      & 2449797054412948224 & 1.8683 & 0.0550 & &    49.753 & 0.089 & & $-$172.407 & 0.051 & & 0.527 & 0.015 & 0.016 & &    $-$9.4 & 0.5 &   \citet{aoki10} \\           
\object[BPS CS 29497-004]{CS~29497-004}                       & 2322729725405593728 & 0.2396 & 0.0299 & &     9.473 & 0.049 & &      0.392 & 0.042 & & 3.474 & 0.319 & 0.384 & &     105.0 & 0.4 &   \citet{hansen15rpro} \\     
\object[BPS CS 31082-001]{CS~31082-001}                       & 2451773941958712192 & 0.4806 & 0.0462 & &    11.745 & 0.093 & &  $-$42.709 & 0.038 & & 1.913 & 0.155 & 0.183 & &     139.1 & 0.1 &   \citet{hansen15rpro} \\     
\object[HD 222925]{HD~222925}                                 & 6487799171512458624 & 2.2332 & 0.0243 & &   154.854 & 0.041 & &  $-$99.171 & 0.041 & & 0.442 & 0.005 & 0.005 & &   $-$38.9 & 0.6 &   \citet{roederer18d} \\      
\object[2MASS J21064294-6828266]{J21064294$-$6828266}         & 6376678403241698560 & 0.4489 & 0.0297 & &  $-$4.429 & 0.034 & &  $-$18.376 & 0.045 & & 2.089 & 0.124 & 0.141 & &   $-$72.6 & 0.6 &   \citet{hansen18} \\         
\object[2MASS J09544277+5246414]{J09544277$+$5246414}         & 828438619475671936  & 0.3061 & 0.0357 & & $-$17.897 & 0.050 & &  $-$26.914 & 0.049 & & 2.809 & 0.251 & 0.301 & &   $-$67.7 & 1.0 &   \citet{holmbeck18rpro} \\       
\object[2MASS J15383085-1804242]{J15383085$-$1804242}         & 6255142030043852928 & 0.9629 & 0.0399 & & $-$49.471 & 0.091 & &  $-$37.986 & 0.060 & & 1.010 & 0.040 & 0.043 & &     131.3 & 0.5 &   \citet{sakari18a} \\        
\object[2MASS J21091825-1310062]{J21091825$-$1310062}         & 6885782695269539584 & 0.4254 & 0.0429 & &     3.183 & 0.069 & &  $-$28.673 & 0.052 & & 2.190 & 0.193 & 0.233 & &   $-$35.9 & 0.4 &   \citet{hansen18} \\         
\object[BPS CS 31078-018]{CS~31078-018}                       & 7189878332862720    & 0.6440 & 0.0300 & &    17.350 & 0.057 & &  $-$10.216 & 0.048 & & 1.479 & 0.063 & 0.069 & &      81.3 & 0.2 &   \citet{lai08} \\            
\object[HE 0430-4901]{HE~0430$-$4901}                         & 4787830774791048832 & 0.3698 & 0.0170 & &     7.281 & 0.029 & &      4.822 & 0.037 & & 2.491 & 0.101 & 0.110 & &     208.7 & 3.0 &   \citet{barklem05heres} \\   
\object[BPS CS 22945-058]{CS~22945-058}                       & 6389179335052544256 & 0.5849 & 0.0250 & &    30.739 & 0.037 & &  $-$22.226 & 0.036 & & 1.627 & 0.064 & 0.069 & &      23.4 & 0.5 &   \citet{roederer14c} \\      
\object[BPS CS 22945-017]{CS~22945-017}                       & 6486692891016362240 & 0.7907 & 0.0212 & &    29.925 & 0.031 & &   $-$3.596 & 0.031 & & 1.220 & 0.031 & 0.033 & &     101.8 & 0.8 &   \citet{roederer14c} \\      
\object[2MASS J02165716-7547064]{J02165716$-$7547064}         & 4637170571951777280 & 0.1824 & 0.0215 & &  $-$3.367 & 0.038 & &      1.648 & 0.032 & & 4.474 & 0.381 & 0.453 & &    $-$5.8 & 0.3 &   \citet{hansen18} \\         
\object[SMSS J062609.83-590503.2]{SMSS~J062609.83$-$590503.2} & 5482786685494509056 & 0.1907 & 0.0169 & &     2.073 & 0.034 & &      2.140 & 0.035 & & 4.402 & 0.302 & 0.347 & &  $-$110.0 & 1.0 &   \citet{jacobson15smss} \\   
\object[HE 2224+0143]{HE~2224$+$0143}                         & 2703430605705583360 & 0.3207 & 0.0369 & &     4.073 & 0.043 & &  $-$10.241 & 0.043 & & 2.762 & 0.256 & 0.310 & &  $-$113.1 & 0.2 &   \citet{hansen15rpro} \\     
\object[BPS CS 22953-003]{CS~22953-003}                       & 4710594687144052096 & 0.2101 & 0.0141 & &  $-$3.819 & 0.029 & &  $-$15.767 & 0.022 & & 4.101 & 0.224 & 0.250 & &     208.5 & 0.3 &   \citet{bonifacio09,roederer14c} \\
\object[BPS CS 22958-052]{CS~22958-052}                       & 4742398404575483648 & 0.8491 & 0.0163 & &    37.371 & 0.028 & &      9.317 & 0.028 & & 1.139 & 0.021 & 0.022 & &      88.9 & 0.8 &   \citet{roederer14c} \\      
\object[SMSS J024858.41-684306.4]{SMSS~J024858.41$-$684306.4} & 4647065936083474816 & 0.2687 & 0.0157 & &  $-$5.104 & 0.030 & &   $-$8.605 & 0.027 & & 3.318 & 0.166 & 0.183 & &  $-$239.2 & 1.0 &   \citet{jacobson15smss} \\   
\object[HE 2327-5642]{HE~2327$-$5642}                         & 6495850379767072128 & 0.1779 & 0.0201 & &    12.569 & 0.026 & &   $-$8.692 & 0.026 & & 4.595 & 0.379 & 0.448 & &     282.2 & 1.0 &   \citet{mashonkina10}\tablenotemark{a} \\
\object[BPS CS 29491-069]{CS~29491-069}                       & 6600971319243174144 & 0.3857 & 0.0326 & &    10.055 & 0.046 & &  $-$34.609 & 0.049 & & 2.386 & 0.174 & 0.202 & &  $-$377.1 & 0.7 &   \citet{hayek09} \\          
\object[SMSS J051008.62-372019.8]{SMSS~J051008.62$-$372019.8} & 4820909925710430976 & 0.9051 & 0.0291 & &    11.816 & 0.038 & &  $-$23.439 & 0.046 & & 1.070 & 0.033 & 0.035 & &     372.8 & 1.0 &   \citet{jacobson15smss} \\   
\object[BD +173248]{BD $+$17\degree3248}                      & 4553184509407224576 & 1.2209 & 0.0359 & & $-$47.746 & 0.064 & &  $-$22.412 & 0.067 & & 0.801 & 0.023 & 0.024 & &  $-$145.2 & 0.8 &   \citet{behr03} \\           
\object[BPS CS 29529-054]{CS~29529-054}                       & 4679456071169507712 & 0.9797 & 0.0199 & &    28.701 & 0.032 & &  $-$28.615 & 0.037 & & 0.991 & 0.019 & 0.020 & &     113.2 & 0.4 &   \citet{roederer14c} \\      
\object[2MASS J15582962-1224344]{J15582962$-$1224344}         & 4342895871148449152 & 0.4103 & 0.0335 & &  $-$8.824 & 0.069 & &   $-$0.691 & 0.041 & & 2.284 & 0.166 & 0.193 & &      83.1 & 0.5 &   \citet{hansen18} \\         
\object[2MASS J00405260-5122491]{J00405260$-$5122491}         & 4925248047268557056 & 6.6710 & 0.0274 & &   211.493 & 0.036 & & $-$197.324 & 0.037 & & 0.149 & 0.001 & 0.001 & &     123.1 & 0.2 &   \citet{hansen18} \\         
\object[BPS CS 22943-132]{CS~22943-132}                       & 6679228303437917696 & 2.1670 & 0.0242 & &    68.310 & 0.039 & &  $-$43.701 & 0.026 & & 0.456 & 0.005 & 0.005 & &      18.9 & 0.8 &   \citet{roederer14c} \\      
\object[BPS CS 22896-154]{CS~22896-154}                       & 6448440159932433536 & 0.3074 & 0.0181 & &  $-$9.747 & 0.028 & &  $-$24.965 & 0.024 & & 2.971 & 0.156 & 0.172 & &     137.9 & 1.0 &   \citet{bonifacio09} \\      
\object[HD 115444]{HD~115444}                                 & 1474455748663044736 & 1.1989 & 0.0403 & &     4.556 & 0.051 & &  $-$60.449 & 0.054 & & 0.815 & 0.026 & 0.028 & &   $-$27.0 & 0.5 &   \citet{hansen15rpro} \\     
\object[HD 20]{HD~20}                                         & 2333756864959936000 & 1.9447 & 0.0527 & &   132.434 & 0.066 & &  $-$39.917 & 0.058 & & 0.507 & 0.013 & 0.014 & &   $-$57.9 & 0.1 &   \citet{hansen15rpro} \\     
\object[HD 221170]{HD~221170}                                 & 2869759781250083200 & 1.8370 & 0.0587 & & $-$16.642 & 0.080 & &  $-$53.664 & 0.059 & & 0.536 & 0.017 & 0.018 & &  $-$121.2 & 0.1 &   \citet{hansen15rpro} \\     
\object[HE 0420+0123a]{HE~0420$+$0123a}                       & 3279770347306973056 & 0.8934 & 0.0427 & &    33.257 & 0.076 & &  $-$21.424 & 0.044 & & 1.083 & 0.048 & 0.053 & &   $-$55.3 & 1.0 &   \citet{hollek11} \\         
\object[2MASS J19232518-5833410]{J19232518$-$5833410}         & 6638565923901510656 & 0.4567 & 0.0548 & & $-$23.412 & 0.072 & &  $-$16.432 & 0.048 & & 2.072 & 0.218 & 0.275 & &     125.9 & 0.6 &   \citet{hansen18} \\         
\object[HE 1430+0053]{HE~1430$+$0053}                         & 3653467682134558592 & 0.2980 & 0.0304 & & $-$23.168 & 0.053 & &  $-$14.932 & 0.044 & & 3.009 & 0.253 & 0.302 & &  $-$107.7 & 0.4 &   \citet{hansen15rpro} \\     
\object[2MASS J01530024-3417360]{J01530024$-$3417360}         & 5017240268153817600 & 3.2385 & 0.0386 & &    22.189 & 0.041 & & $-$212.060 & 0.045 & & 0.306 & 0.004 & 0.004 & &      22.8 & 0.4 &   \citet{hansen18} \\         
\object[2MASS J15271353-2336177]{J15271353$-$2336177}         & 6239162964995926016 & 6.7911 & 0.0456 & & $-$228.893& 0.095 & &  $-$96.951 & 0.066 & & 0.147 & 0.001 & 0.001 & &       1.2 & 0.3 &   \citet{hansen18} \\         
\enddata
\tablecomments{The $-$ and $+$ columns indicate the 
16th percentile and 84th percentile confidence intervals}
\tablenotetext{a}{The median RV from this source is adopted as the systemic RV}
\end{deluxetable*}

\begin{deluxetable*}{lDDDcDDDcDDDcDDD}
\tablecaption{Calculated Velocities in a Cylindrical Coordinate System
\label{cylindricalvelocitytab}}
\tablewidth{0pt}
\tabletypesize{\scriptsize}
\tablehead{
\colhead{Star} &
\twocolhead{$V_{R}$} &
\twocolhead{$-$} &
\twocolhead{$+$} &
\colhead{} &
\twocolhead{$V_{\phi}$} &
\twocolhead{$-$} &
\twocolhead{$+$} &
\colhead{} &
\twocolhead{$V_{z}$} &
\twocolhead{$-$} &
\twocolhead{$+$} &
\colhead{} &
\twocolhead{$V_{\perp}$} &
\twocolhead{$-$} &
\twocolhead{$+$} \\
\cline{2-7}
\cline{9-14}
\cline{16-21}
\cline{23-28} 
\colhead{} &
\multicolumn{6}{c}{(km s$^{-1}$)} &
\colhead{} &
\multicolumn{6}{c}{(km s$^{-1}$)} &
\colhead{} &
\multicolumn{6}{c}{(km s$^{-1}$)} &
\colhead{} &
\multicolumn{6}{c}{(km s$^{-1}$)} 
}
\decimals
\startdata
\object[SDSS J235718.91-005247.8]{J235718.91$-$005247.8}       & $-$118.3 &  5.7 &   5.6 & & $-$154.4 & 18.4 & 17.6 & & $-$204.7 & 10.6 & 10.2 & & 236.4 & 11.5 & 12.1  \\
\object[BPS CS 29497-004]{CS~29497-004}                        &    153.2 & 23.7 &  37.2 & &    145.7 & 21.8 & 13.4 & & $-$114.2 &  3.9 &  2.5 & & 191.0 & 20.1 & 33.1  \\
\object[BPS CS 31082-001]{CS~31082-001}                        & $-$130.6 & 14.0 &  12.2 & & $-$161.6 & 34.5 & 30.2 & & $-$193.3 &  5.7 &  5.1 & & 233.3 & 10.9 & 12.7  \\
\object[HD 222925]{HD~222925}                                  &    246.9 &  4.3 &   4.8 & &  $-$52.2 &  5.6 &  5.0 & &     68.5 &  0.7 &  0.7 & & 256.2 &  4.3 &  4.8  \\
\object[2MASS J21064294-6828266]{J21064294$-$6828266}          &     56.9 &  3.5 &   4.8 & &    100.8 & 15.8 & 12.8 & &    142.9 &  7.4 &  9.0 & & 153.8 &  8.0 & 10.2  \\
\object[2MASS J09544277+5246414]{J09544277$+$5246414}          &    161.9 & 23.3 &  25.8 & & $-$237.0 & 65.0 & 55.3 & & $-$119.4 & 11.0 &  9.4 & & 201.2 & 23.8 & 27.6  \\
\object[2MASS J15383085-1804242]{J15383085$-$1804242}          &  $-$52.3 &  3.2 &   3.5 & &  $-$88.0 & 13.2 & 11.8 & &     96.3 &  1.0 &  1.1 & & 109.6 &  0.7 &  0.8  \\
\object[2MASS J21091825-1310062]{J21091825$-$1310062}          &  $-$84.4 & 14.6 &  11.7 & &  $-$52.2 & 24.0 & 20.8 & & $-$102.8 & 13.2 & 11.1 & & 133.2 & 16.0 & 19.3  \\
\object[BPS CS 31078-018]{CS~31078-018}                        &     91.2 &  2.4 &   2.7 & &     99.3 &  9.1 &  7.9 & &  $-$28.8 &  1.2 &  1.4 & &  95.6 &  1.9 &  2.2  \\
\object[HE 0430-4901]{HE~0430$-$4901}                          &     98.4 &  7.6 &   9.7 & &     68.8 &  2.0 &  2.1 & &  $-$71.4 &  6.3 &  7.9 & & 121.9 &  3.1 &  4.2  \\
\object[BPS CS 22945-058]{CS~22945-058}                        &    158.5 & 12.2 &  15.0 & &   $-$6.1 & 14.5 & 12.0 & &     41.4 &  3.0 &  3.7 & & 163.9 & 12.7 & 15.4  \\
\object[BPS CS 22945-017]{CS~22945-017}                        &     85.9 &  7.3 &   8.3 & &    114.8 &  3.4 &  3.1 & & $-$121.9 &  2.8 &  2.5 & & 149.0 &  5.9 &  7.3  \\
\object[2MASS J02165716-7547064]{J02165716$-$7547064}          & $-$194.0 & 56.5 &  34.2 & &    240.9 & 25.4 &  9.2 & &  $-$46.5 & 16.7 & 10.3 & & 199.5 & 35.7 & 59.0  \\
\object[SMSS J062609.83-590503.2]{SMSS~J062609.83$-$590503.2}  & $-$126.1 & 15.5 &  12.4 & &    280.4 & 13.3 &  9.2 & &    109.0 &  9.2 & 15.2 & & 166.9 & 15.4 & 21.8  \\
\object[HE 2224+0143]{HE~2224$+$0143}                          &      5.0 &  6.7 &   4.7 & &     36.1 & 16.0 & 11.9 & &  $-$17.0 & 15.1 & 11.0 & &  18.1 &  6.2 & 14.0  \\
\object[BPS CS 22953-003]{CS~22953-003}                        & $-$274.9 & 17.9 &  18.2 & & $-$154.7 & 87.0 & 51.7 & &     31.1 & 28.8 & 45.4 & & 277.1 & 19.9 & 26.7  \\
\object[BPS CS 22958-052]{CS~22958-052}                        &    155.0 &  8.3 &   8.6 & &     92.3 &  3.7 &  3.7 & &  $-$26.1 &  1.8 &  2.1 & & 157.2 &  7.8 &  8.2  \\
\object[SMSS J024858.41-684306.4]{SMSS~J024858.41$-$684306.4}  & $-$255.4 & 45.5 &  34.9 & &    336.1 & 24.6 & 16.2 & &    222.0 &  5.8 &  7.4 & & 338.3 & 29.6 & 40.4  \\
\object[HE 2327-5642]{HE~2327$-$5642}                          &    156.9 & 67.4 & 125.0 & & $-$138.9 & 31.3 & 36.7 & & $-$230.3 &  1.3 &  1.2 & & 279.1 & 32.2 & 85.4  \\
\object[BPS CS 29491-069]{CS~29491-069}                        &    157.0 &  4.7 &   3.6 & & $-$262.2 & 47.9 & 40.3 & &    271.5 &  6.5 &  5.8 & & 313.6 &  7.9 &  6.8  \\
\object[SMSS J051008.62-372019.8]{SMSS~J051008.62$-$372019.8}  &     43.6 &  4.1 &   3.5 & & $-$114.4 &  4.2 &  3.6 & & $-$175.1 &  1.4 &  1.6 & & 180.5 &  2.4 &  2.2  \\
\object[BD +173248]{BD $+$17\degree3248}                       &     37.5 &  2.1 &   1.9 & &   $-$8.6 &  4.9 &  4.7 & &     68.1 &  3.9 &  3.9 & &  77.8 &  2.4 &  2.6  \\
\object[BPS CS 29529-054]{CS~29529-054}                        &  $-$93.1 &  2.9 &   2.5 & &     20.0 &  6.2 &  5.2 & &     64.4 &  5.1 &  6.1 & & 113.1 &  4.8 &  6.0  \\
\object[2MASS J15582962-1224344]{J15582962$-$1224344}          &  $-$47.2 &  3.2 &   4.1 & &    156.6 &  8.2 &  6.3 & &    110.4 &  5.6 &  6.8 & & 120.0 &  3.8 &  5.0  \\
\object[2MASS J00405260-5122491]{J00405260$-$5122491}          &     20.9 &  0.4 &   0.4 & &      2.8 &  1.1 &  1.2 & &  $-$53.6 &  0.4 &  0.4 & &  57.5 &  0.2 &  0.2  \\
\object[BPS CS 22943-132]{CS~22943-132}                        &     66.4 &  1.7 &   1.9 & &    158.3 &  1.4 &  1.2 & & $-$135.0 &  2.5 &  2.4 & & 150.4 &  2.8 &  3.0  \\
\object[BPS CS 22896-154]{CS~22896-154}                        &   $-$6.3 & 21.1 &  32.9 & & $-$209.3 & 52.7 & 42.8 & &     38.6 & 10.7 & 13.9 & &  41.6 &  3.6 & 17.3  \\
\object[HD 115444]{HD~115444}                                  & $-$166.1 &  5.2 &   5.2 & &     52.0 &  5.8 &  5.4 & &     16.6 &  1.2 &  1.3 & & 166.9 &  5.3 &  5.3  \\
\object[HD 20]{HD~20}                                          &    234.0 &  4.7 &   4.9 & &   $-$7.4 &  4.9 &  4.6 & &      6.2 &  1.2 &  1.1 & & 234.1 &  4.7 &  4.9  \\
\object[HD 221170]{HD~221170}                                  & $-$128.3 &  1.9 &   2.1 & &    102.0 &  0.5 &  0.4 & &  $-$33.2 &  2.0 &  2.1 & & 132.5 &  2.5 &  2.4  \\
\object[HE 0420+0123a]{HE~0420$+$0123a}                        &  $-$56.3 &  0.9 &   0.8 & &     44.5 &  9.5 &  8.2 & &    107.8 &  2.9 &  3.5 & & 121.6 &  2.7 &  3.3  \\
\object[2MASS J19232518-5833410]{J19232518$-$5833410}          & $-$125.6 &  1.6 &   2.4 & &  $-$50.1 & 22.5 & 20.5 & &    141.6 & 16.0 & 17.6 & & 189.2 & 10.5 & 12.1  \\
\object[HE 1430+0053]{HE~1430$+$0053}                          &    183.7 & 18.5 &  25.0 & & $-$164.3 & 60.1 & 48.1 & &  $-$45.1 &  4.0 &  5.3 & & 189.2 & 16.7 & 23.3  \\
\object[2MASS J01530024-3417360]{J01530024$-$3417360}          & $-$189.0 &  2.1 &   2.2 & &  $-$26.0 &  3.1 &  3.2 & &     26.0 &  0.6 &  0.6 & & 190.8 &  2.3 &  2.2  \\
\object[2MASS J15271353-2336177]{J15271353$-$2336177}          &     56.4 &  0.5 &   0.4 & &     78.8 &  1.0 &  0.8 & &     50.1 &  0.3 &  0.3 & &  75.4 &  0.4 &  0.5  \\
\enddata
\tablecomments{The $-$ and $+$ columns indicate the 
16th percentile and 84th percentile confidence intervals}
\end{deluxetable*}

\begin{deluxetable*}{lDDDcDDDcDDDcDDD}
\tablecaption{Calculated Orbital Energies and Angular Momenta
\label{kinematictab}}
\tablewidth{0pt}
\tabletypesize{\scriptsize}
\tablehead{
\colhead{Star} &
\twocolhead{$E$} &
\twocolhead{$-$} &
\twocolhead{$+$} &
\colhead{} &
\twocolhead{$J_{r}$} &
\twocolhead{$-$} &
\twocolhead{$+$} &
\colhead{} &
\twocolhead{$J_{\phi}$} &
\twocolhead{$-$} &
\twocolhead{$+$} &
\colhead{} &
\twocolhead{$J_{z}$} &
\twocolhead{$-$} &
\twocolhead{$+$} \\
\cline{2-7}
\cline{9-14}
\cline{16-21}
\cline{23-28} 
\colhead{} &
\multicolumn{6}{c}{($10^{3}$ km$^{2}$ s$^{-2}$)} &
\colhead{} &
\multicolumn{6}{c}{(kpc km s$^{-1}$)} &
\colhead{} &
\multicolumn{6}{c}{(kpc km s$^{-1}$)} &
\colhead{} &
\multicolumn{6}{c}{(kpc km s$^{-1}$)} 
}
\decimals
\startdata
\object[SDSS J235718.91-005247.8]{J235718.91$-$005247.8}       &  $-$91.8 &  5.2 &  5.9 & &   264.  &   80.4 &   142.  & & $-$1210   & 148.  & 141.  & &  760.  &  27.7 &  28.2 \\
\object[BPS CS 29497-004]{CS~29497-004}                        &  $-$94.0 &  3.1 &  6.4 & &   608.  &  176.  &   352.  & &    1150   & 176.  & 110.  & &  254.  &  14.2 &  41.7 \\
\object[BPS CS 31082-001]{CS~31082-001}                        &  $-$86.1 &  7.3 &  9.6 & &   291.  &  146.  &   333.  & & $-$1340   & 302.  & 261.  & &  947.  &   4.8 &  13.3 \\
\object[HD 222925]{HD~222925}                                  &  $-$98.2 &  1.3 &  1.5 & &  1080   &   18.0 &    23.6 & &  $-$378.  &  43.8 &  38.9 & &   90.4 &   2.6 &   2.8 \\  
\object[2MASS J21064294-6828266]{J21064294$-$6828266}          & $-$120.  &  0.3 &  0.6 & &    58.3 &    3.9 &     2.1 & &     692.  & 116.  &  96.5 & &  420.  &  64.5 &  93.1 \\
\object[2MASS J09544277+5246414]{J09544277$+$5246414}          &  $-$70.1 & 17.4 & 25.0 & &   882.  &  592.  &  2600   & & $-$2360   & 745.  & 605.  & &  527.  &  61.5 &  86.8 \\
\object[2MASS J15383085-1804242]{J15383085$-$1804242}          & $-$127.  &  0.7 &  0.9 & &   222.  &   41.5 &    38.4 & &  $-$599.  &  90.0 &  81.0 & &  146.  &   1.3 &   1.4 \\
\object[2MASS J21091825-1310062]{J21091825$-$1310062}          & $-$128.  &  2.4 &  3.8 & &   265.  &   87.9 &    93.8 & &  $-$320.  & 149.  & 133.  & &  293.  &  68.4 &  80.6 \\
\object[BPS CS 31078-018]{CS~31078-018}                        & $-$114.  &  0.2 &  0.3 & &   375.  &   37.0 &    45.0 & &     938.  &  76.1 &  65.2 & &   26.4 &   0.6 &   0.8 \\
\object[HE 0430-4901]{HE~0430$-$4901}                          & $-$115.  &  1.0 &  1.4 & &   473.  &   32.1 &    40.6 & &     634.  &  20.0 &  20.0 & &  106.  &  14.1 &  15.4 \\ 
\object[BPS CS 22945-058]{CS~22945-058}                        & $-$120.  &  1.9 &  2.7 & &   704.  &   25.8 &    21.7 & &   $-$16.5 & 105.  &  87.7 & &  106.  &  18.8 &  26.5 \\
\object[BPS CS 22945-017]{CS~22945-017}                        & $-$115.  &  0.4 &  0.6 & &   197.  &   21.2 &    25.5 & &     880.  &  28.8 &  26.8 & &  220.  &   8.4 &   9.4 \\
\object[2MASS J02165716-7547064]{J02165716$-$7547064}          &  $-$77.8 &  6.3 & 11.8 & &   827.  &  265.  &   677.  & &    1760   & 142.  &  59.8 & &  445.  & 164.  & 407.  \\
\object[SMSS J062609.83-590503.2]{SMSS~J062609.83$-$590503.2}  &  $-$66.6 &  2.3 &  4.2 & &  1230   &  148.  &   297.  & &    2670   &  32.4 &  77.0 & &  183.  &  34.5 &  62.8 \\
\object[HE 2224+0143]{HE~2224$+$0143}                          & $-$130.  &  0.1 &  0.5 & &   439.  &   37.5 &    44.0 & &     293.  & 120.  &  90.8 & &   77.0 &  14.6 &  30.4 \\
\object[BPS CS 22953-003]{CS~22953-003}                        &  $-$77.5 & 13.3 & 27.0 & &  1350   &  410.  &  2210   & & $-$1060   & 588.  & 354.  & &  404.  &  45.3 & 155.  \\
\object[BPS CS 22958-052]{CS~22958-052}                        & $-$114.  &  0.9 &  1.0 & &   510.  &   38.1 &    41.3 & &     759.  &  29.3 &  29.3 & &   22.6 &   0.7 &   1.0 \\
\object[SMSS J024858.41-684306.4]{SMSS~J024858.41$-$684306.4}  &  $-$13.4 &  5.2 &  7.7 & & 24200   & 7770   & 31300   & &    2580   & 181.  & 123.  & &  343.  &   7.4 &   4.4 \\
\object[HE 2327-5642]{HE~2327$-$5642}                          &  $-$80.6 & 14.8 & 37.0 & &  1260   &  713.  &  3930   & &  $-$788.  & 117.  & 182.  & &  592.  & 110.  & 273.  \\
\object[BPS CS 29491-069]{CS~29491-069}                        &  $-$51.5 &  7.6 & 11.1 & &  2970   &  916.  &  2030   & & $-$1740   & 281.  & 246.  & & 1080   &  89.4 & 105.  \\
\object[SMSS J051008.62-372019.8]{SMSS~J051008.62$-$372019.8}  & $-$106.  &  0.2 &  0.3 & &    59.9 &    7.7 &     6.4 & &  $-$936.  &  38.5 &  32.3 & &  650.  &  17.4 &  15.6 \\
\object[BD +173248]{BD $+$17\degree3248}                       & $-$132.  &  0.1 &  0.1 & &   566.  &   31.9 &    20.5 & &   $-$36.3 &  36.2 &  35.0 & &   57.8 &   8.1 &   9.1 \\
\object[BPS CS 29529-054]{CS~29529-054}                        & $-$124.  &  0.4 &  0.6 & &   596.  &   18.8 &    22.0 & &     190.  &  49.4 &  41.4 & &   58.5 &   9.9 &  13.8 \\
\object[2MASS J15582962-1224344]{J15582962$-$1224344}          & $-$123.  &  1.9 &  1.5 & &    21.3 &    1.7 &     2.0 & &     943.  &  84.0 &  67.4 & &  193.  &  22.4 &  28.9 \\
\object[2MASS J00405260-5122491]{J00405260$-$5122491}          & $-$130.  &  0.1 &  0.1 & &   608.  &    7.3 &     6.2 & &      52.2 &   8.8 &   9.8 & &   33.0 &   0.6 &   0.6 \\
\object[BPS CS 22943-132]{CS~22943-132}                        & $-$109.  &  0.2 &  0.2 & &    73.0 &    3.3 &     3.9 & &    1240   &  11.5 &  10.7 & &  251.  &  10.7 &  11.4 \\
\object[BPS CS 22896-154]{CS~22896-154}                        & $-$124.  &  6.6 & 11.8 & &    25.8 &   22.3 &    79.5 & & $-$1110   & 192.  & 185.  & &   90.1 &  26.2 &  49.6 \\
\object[HD 115444]{HD~115444}                                  & $-$115.  &  0.6 &  0.6 & &   665.  &   34.2 &    36.9 & &     446.  &  46.2 &  43.0 & &   27.7 &   2.6 &   3.1 \\
\object[HD 20]{HD~20}                                          & $-$104.  &  1.1 &  1.2 & &  1210   &   26.7 &    16.5 & &   $-$28.9 &  38.6 &  36.2 & &   10.8 &   0.1 &   0.1 \\
\object[HD 221170]{HD~221170}                                  & $-$116.  &  0.3 &  0.3 & &   393.  &    5.2 &     5.4 & &     858.  &   3.5 &   3.2 & &   15.8 &   1.9 &   2.1 \\
\object[HE 0420+0123a]{HE~0420$+$0123a}                        & $-$117.  &  0.2 &  0.2 & &   387.  &   21.8 &    23.4 & &     432.  &  83.3 &  71.2 & &  285.  &  30.9 &  42.2 \\
\object[2MASS J19232518-5833410]{J19232518$-$5833410}          & $-$123.  &  1.9 &  2.8 & &   397.  &   75.5 &    82.9 & &  $-$288.  & 129.  & 124.  & &  259.  &  58.9 &  64.8 \\
\object[HE 1430+0053]{HE~1430$+$0053}                          & $-$105.  &  9.5 & 16.3 & &   322.  &   38.6 &   275.  & &  $-$975.  & 300.  & 266.  & &  314.  &  52.2 &  96.4 \\
\object[2MASS J01530024-3417360]{J01530024$-$3417360}          & $-$113.  &  0.5 &  0.5 & &   895.  &    3.1 &     4.2 & &  $-$178.  &  24.6 &  25.5 & &    6.1 &   0.2 &   0.2 \\
\object[2MASS J15271353-2336177]{J15271353$-$2336177}          & $-$126.  &  0.1 &  0.1 & &   334.  &    2.8 &     3.5 & &     650.  &   7.6 &   6.3 & &   24.6 &   0.3 &   0.4 \\
\enddata
\tablecomments{The $-$ and $+$ columns indicate the 
16th percentile and 84th percentile confidence intervals}
\end{deluxetable*}

\begin{deluxetable*}{lDDDcDDDcDDDcDDD}
\tablecaption{Calculated Orbital Parameters
\label{orbittab}}
\tablewidth{0pt}
\tabletypesize{\scriptsize}
\tablehead{
\colhead{Star} &
\twocolhead{$r_{\rm peri}$} &
\twocolhead{$-$} &
\twocolhead{$+$} &
\colhead{} &
\twocolhead{$r_{\rm apo}$} &
\twocolhead{$-$} &
\twocolhead{$+$} &
\colhead{} &
\twocolhead{$Z_{\rm max}$} &
\twocolhead{$-$} &
\twocolhead{$+$} &
\colhead{} &
\twocolhead{$e$} &
\twocolhead{$-$} &
\twocolhead{$+$} \\
\cline{2-7}
\cline{9-14}
\cline{16-21}
\cline{23-28}
\colhead{} &
\multicolumn{6}{c}{(kpc)} &
\colhead{} &
\multicolumn{6}{c}{(kpc)} &
\colhead{} &
\multicolumn{6}{c}{(kpc)} &
\colhead{} &
\colhead{} &
\colhead{} &
\colhead{} 
}
\decimals
\startdata
\object[SDSS J235718.91-005247.8]{J235718.91$-$005247.8}       & 6.60 & 0.43 & 0.32 & &  15.9  &   2.41 &   3.40 & &  12.0  &  1.88 &  2.48 & & 0.414 & 0.041 & 0.059 \\
\object[BPS CS 29497-004]{CS~29497-004}                        & 3.71 & 0.71 & 0.56 & &  16.9  &   1.81 &   3.89 & &   8.24 &  1.27 &  2.94 & & 0.641 & 0.081 & 0.109 \\
\object[BPS CS 31082-001]{CS~31082-001}                        & 7.86 & 0.38 & 0.21 & &  18.5  &   4.11 &   7.29 & &  14.6  &  2.68 &  4.80 & & 0.403 & 0.088 & 0.120 \\
\object[HD 222925]{HD~222925}                                  & 0.99 & 0.07 & 0.07 & &  16.6  &   0.52 &   0.64 & &   5.28 &  0.08 &  0.09 & & 0.888 & 0.003 & 0.004 \\  
\object[2MASS J21064294-6828266]{J21064294$-$6828266}          & 3.96 & 0.10 & 0.18 & &   7.19 &   0.06 &   0.06 & &   5.20 &  0.43 &  0.52 & & 0.289 & 0.016 & 0.010 \\
\object[2MASS J09544277+5246414]{J09544277$+$5246414}          & 8.67 & 1.30 & 0.79 & &  30.6  &  13.2  &  43.2  & &  16.9  &  6.96 & 20.1  & & 0.559 & 0.152 & 0.213 \\
\object[2MASS J15383085-1804242]{J15383085$-$1804242}          & 2.02 & 0.32 & 0.39 & &   7.42 &   0.05 &   0.05 & &   3.21 &  0.10 &  0.10 & & 0.571 & 0.065 & 0.058 \\
\object[2MASS J21091825-1310062]{J21091825$-$1310062}          & 1.57 & 0.83 & 0.85 & &   7.28 &   0.12 &   0.22 & &   5.23 &  0.08 &  0.28 & & 0.644 & 0.133 & 0.168 \\
\object[BPS CS 31078-018]{CS~31078-018}                        & 2.56 & 0.27 & 0.24 & &  10.4  &   0.09 &   0.11 & &   1.41 &  0.04 &  0.05 & & 0.605 & 0.032 & 0.037 \\
\object[HE 0430-4901]{HE~0430$-$4901}                          & 1.81 & 0.08 & 0.11 & &  10.5  &   0.37 &   0.44 & &   3.53 &  0.18 &  0.23 & & 0.706 & 0.012 & 0.012 \\ 
\object[BPS CS 22945-058]{CS~22945-058}                        & 0.17 & 0.10 & 0.20 & &   9.72 &   0.22 &   0.52 & &   4.55 &  0.41 &  0.33 & & 0.965 & 0.037 & 0.020 \\
\object[BPS CS 22945-017]{CS~22945-017}                        & 3.30 & 0.14 & 0.14 & &   9.20 &   0.20 &   0.26 & &   4.58 &  0.21 &  0.26 & & 0.472 & 0.025 & 0.027 \\
\object[2MASS J02165716-7547064]{J02165716$-$7547064}          & 6.27 & 0.05 & 0.25 & &  25.9  &   4.80 &  12.1  & &  14.6  &  5.01 & 13.2  & & 0.611 & 0.068 & 0.095 \\
\object[SMSS J062609.83-590503.2]{SMSS~J062609.83$-$590503.2}  & 8.04 & 0.07 & 0.21 & &  36.4  &   2.78 &   5.58 & &  11.6  &  1.85 &  3.64 & & 0.638 & 0.021 & 0.034 \\
\object[HE 2224+0143]{HE~2224$+$0143}                          & 1.10 & 0.35 & 0.03 & &   7.76 &   0.04 &   0.23 & &   2.80 &  0.08 &  0.28 & & 0.752 & 0.008 & 0.077 \\
\object[BPS CS 22953-003]{CS~22953-003}                        & 3.36 & 1.01 & 1.31 & &  26.8  &   8.02 &  33.2  & &  16.2  &  4.39 & 19.6  & & 0.783 & 0.009 & 0.075 \\
\object[BPS CS 22958-052]{CS~22958-052}                        & 1.92 & 0.10 & 0.10 & &  11.0  &   0.31 &   0.34 & &   1.34 &  0.06 &  0.07 & & 0.703 & 0.021 & 0.021 \\
\object[SMSS J024858.41-684306.4]{SMSS~J024858.41$-$684306.4}  & 5.71 & 0.62 & 0.44 & & 434.   & 138.   & 140.   & & 151.   & 53.0  & 60.9  & & 0.974 & 0.015 & 0.008 \\
\object[HE 2327-5642]{HE~2327$-$5642}                          & 3.02 & 0.21 & 0.64 & &  23.0  &   8.07 &  31.8  & &  16.9  &  3.89 & 21.2  & & 0.770 & 0.168 & 0.129 \\
\object[BPS CS 29491-069]{CS~29491-069}                        & 6.89 & 0.03 & 0.07 & &  62.9  &  15.2  &  34.6  & &  48.5  &  9.59 & 22.1  & & 0.803 & 0.054 & 0.064 \\
\object[SMSS J051008.62-372019.8]{SMSS~J051008.62$-$372019.8}  & 6.04 & 0.14 & 0.18 & &   9.93 &   0.07 &   0.07 & &   7.71 &  0.17 &  0.16 & & 0.244 & 0.017 & 0.013 \\
\object[BD +173248]{BD $+$17\degree3248}                       & 0.13 & 0.08 & 0.11 & &   7.74 &   0.00 &   0.00 & &   3.43 &  2.06 &  0.09 & & 0.967 & 0.027 & 0.021 \\
\object[BPS CS 29529-054]{CS~29529-054}                        & 0.56 & 0.03 & 0.02 & &   8.96 &   0.08 &   0.13 & &   3.63 &  1.87 &  0.04 & & 0.884 & 0.006 & 0.005 \\
\object[2MASS J15582962-1224344]{J15582962$-$1224344}          & 4.47 & 0.21 & 0.16 & &   6.31 &   0.28 &   0.26 & &   3.02 &  0.14 &  0.17 & & 0.173 & 0.004 & 0.004 \\
\object[2MASS J00405260-5122491]{J00405260$-$5122491}          & 0.13 & 0.03 & 0.02 & &   8.12 &   0.01 &   0.08 & &   1.05 &  0.01 &  0.02 & & 0.969 & 0.004 & 0.007 \\
\object[BPS CS 22943-132]{CS~22943-132}                        & 5.43 & 0.03 & 0.03 & &   9.41 &   0.07 &   0.08 & &   4.65 &  0.14 &  0.15 & & 0.268 & 0.006 & 0.007 \\
\object[BPS CS 22896-154]{CS~22896-154}                        & 4.93 & 1.41 & 0.48 & &   6.05 &   0.21 &   2.76 & &   1.78 &  0.25 &  1.40 & & 0.184 & 0.114 & 0.172 \\
\object[HD 115444]{HD~115444}                                  & 1.05 & 0.12 & 0.11 & &  11.0  &   0.18 &   0.20 & &   1.43 &  0.10 &  0.12 & & 0.826 & 0.020 & 0.020 \\
\object[HD 20]{HD~20}                                          & 0.19 & 0.02 & 0.04 & &  14.7  &   0.38 &   0.43 & &   1.34 &  0.02 &  0.01 & & 0.975 & 0.004 & 0.002 \\
\object[HD 221170]{HD~221170}                                  & 2.26 & 0.01 & 0.02 & &  10.1  &   0.08 &   0.06 & &   0.93 &  0.07 &  0.10 & & 0.634 & 0.005 & 0.003 \\
\object[HE 0420+0123a]{HE~0420$+$0123a}                        & 1.61 & 0.13 & 0.18 & &   9.57 &   0.06 &   0.10 & &   6.10 &  0.51 &  0.57 & & 0.713 & 0.030 & 0.021 \\
\object[2MASS J19232518-5833410]{J19232518$-$5833410}          & 1.04 & 0.52 & 0.66 & &   8.57 &   0.14 &   0.29 & &   5.40 &  0.67 &  0.62 & & 0.784 & 0.105 & 0.100 \\
\object[HE 1430+0053]{HE~1430$+$0053}                          & 3.73 & 1.11 & 0.93 & &  12.0  &   2.26 &   6.89 & &   6.86 &  1.36 &  4.39 & & 0.555 & 0.024 & 0.088 \\
\object[2MASS J01530024-3417360]{J01530024$-$3417360}          & 0.38 & 0.06 & 0.05 & &  11.9  &   0.12 &   0.13 & &   0.69 &  0.02 &  0.02 & & 0.938 & 0.008 & 0.008 \\
\object[2MASS J15271353-2336177]{J15271353$-$2336177}          & 1.75 & 0.02 & 0.02 & &   8.23 &   0.00 &   0.00 & &   1.08 &  0.01 &  0.01 & & 0.648 & 0.003 & 0.004 \\
\enddata
\tablecomments{The $-$ and $+$ columns indicate the 
16th percentile and 84th percentile confidence intervals}
\end{deluxetable*}

\section{Discussion}
\label{discussion}

\subsection{Kinematic and Orbital Properties of
Highly {\it R}-Process-Enhanced Stars in the Solar Neighborhood}
\label{orbitresults}

\begin{figure*}
\begin{center}
\includegraphics[angle=0,width=6.5in]{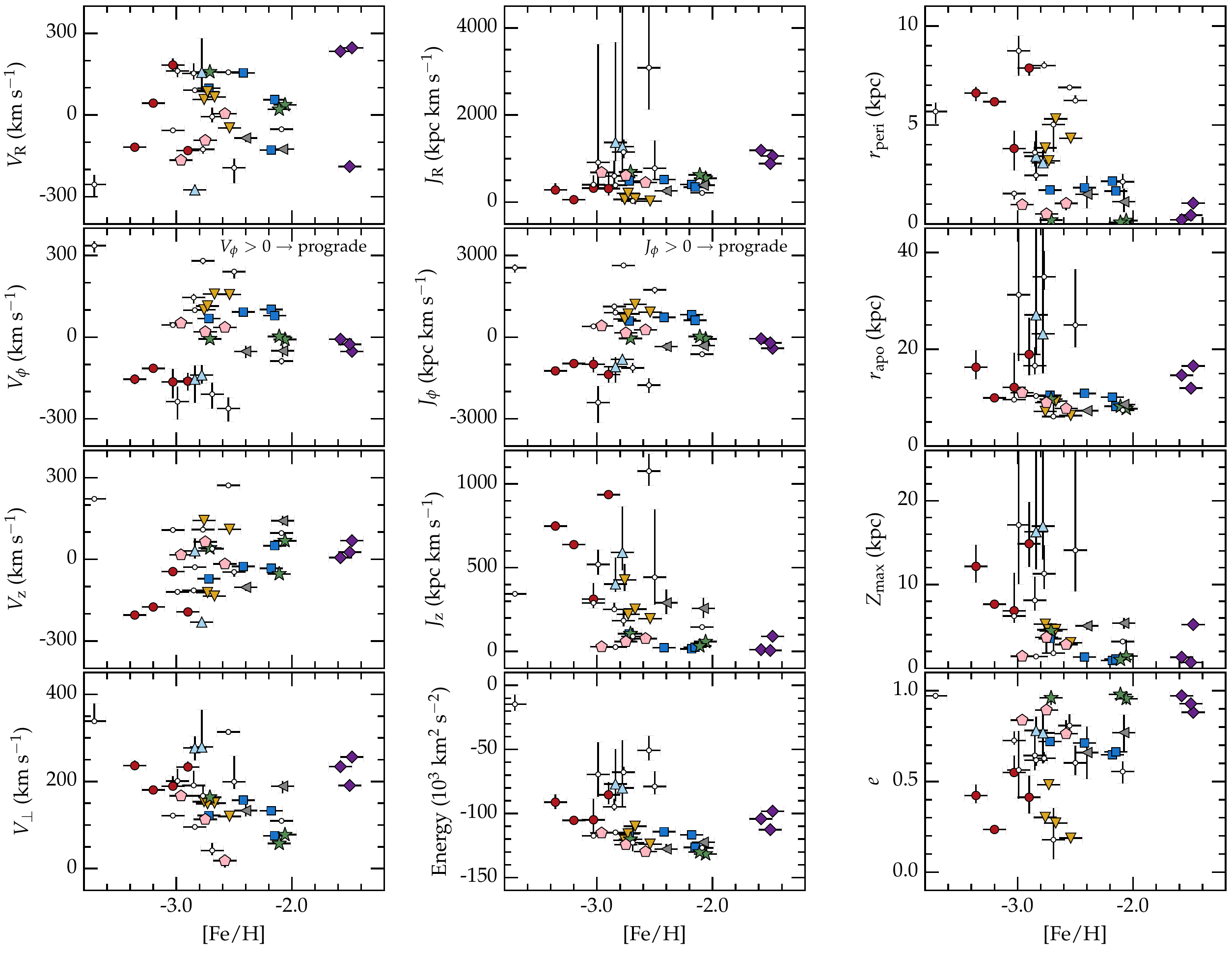} 
\end{center}
\caption{
\label{fehgroupplot}
Calculated orbital and kinematic properties of 
groups of field \rpro-enhanced stars
as functions of [Fe/H].
Each group of stars is indicated by a different symbol/color
combination, as shown in Figure~\ref{energyactiongroupplot}
and discussed in Section~\ref{clustering}.
The error bars on the orbital and kinematic quantities
represent the 16$^{\rm th}$ and 84$^{\rm th}$
percentiles of the distributions, and
the error bars on [Fe/H] show a representative
typical 0.10~dex uncertainty.
 }
\end{figure*}

In this section,
we discuss general kinematic and orbital properties of the ensemble of 
highly \rpro-enhanced stars and
highlight characteristics of a few individual stars.
Figure~\ref{fehgroupplot}
shows the relationships between
[Fe/H] and our calculated
orbital and kinematic properties.
Different symbol shapes and colors are used 
in Figure~\ref{fehgroupplot}
to indicate the different groups of stars
identified by our clustering analysis (Section~\ref{clustering}).
For now, we disregard these classifications.

\begin{figure}
\begin{center}
\includegraphics[angle=0,width=3.35in]{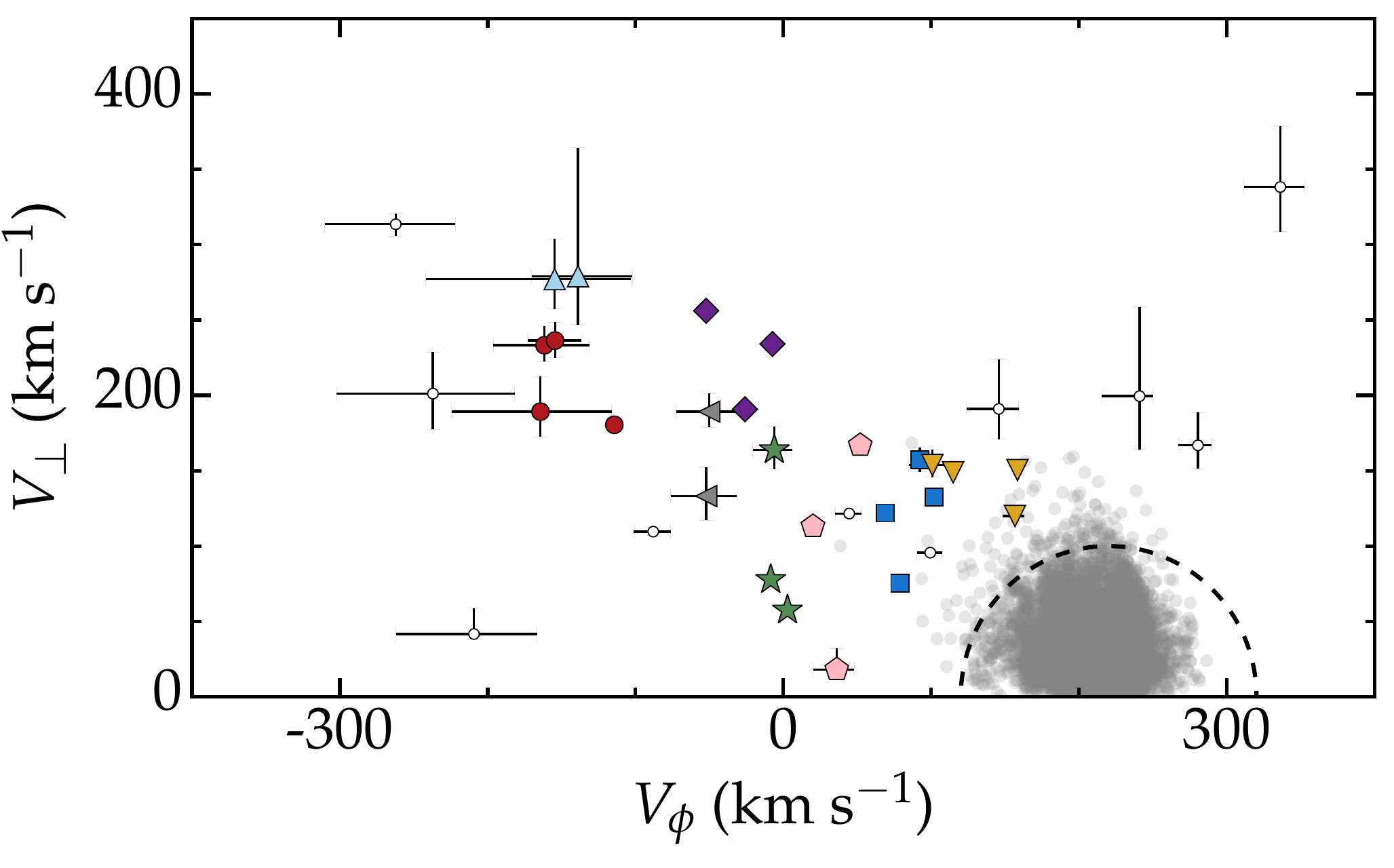}
\end{center}
\caption{
\label{toomreplot}
Toomre diagram in cylindrical coordinates for
our sample of 35 highly \rpro-enhanced stars.
The cloud of gray circles centered near $V_{\phi} \approx$~220~\kmsec\
is a collection of 10,385 disk stars 
located within 200~pc of the Sun
with $-$0.25~$<$~[Fe/H]~$< +$0.25.
The dashed line represents a total space velocity of
100~\kmsec\ relative to the 
local standard of rest in the $V_{\phi}$ direction.
Each group of stars is indicated by a different symbol/color
combination, as shown in Figure~\ref{energyactiongroupplot}
and discussed in Section~\ref{clustering}.
}
\end{figure}

None of the stars in our sample have disk-like kinematics,
as shown in Figure~\ref{toomreplot}.
For comparison, Figure~\ref{toomreplot} also shows
a sample of 10,385 disk stars 
located within 200~pc of the Sun 
with $-$0.25~$<$~[Fe/H]~$< +$0.25
(cf.\ \citealt{hattori18hercules}),
selected from the 
Tycho-\textit{Gaia} Astrometric Solution 
(TGAS; \citealt{lindegren16}) and 
RAdial Velocity Experiment (RAVE DR5; \citealt{kunder17}).
There is virtually no velocity overlap between these disk stars
and the highly \rpro-enhanced stars.
If the stars in our sample are not on disk-like orbits,
then presumably they must have been 
formed in situ in the halo, 
formed in the disk or bulge
and ejected into halo orbits, or
accreted.
We argue in Section~\ref{environment}
that the accretion origin is
likely for most of these stars.

All of the highly \rpro-enhanced stars in our sample
are bound to the Milky Way 
($E <$~0; see Figure~\ref{fehgroupplot}).
There are roughly equal numbers of stars moving 
toward 
($V_{\rm R} <$~0) and 
away from the Galactic center
($V_{\rm R} >$~0), 
and 
there are roughly equal numbers of stars
moving north
($V_{z} >$~0)
and south
($V_{z} <$~0)
as they pass through the Galactic disk.
There are also roughly equal numbers of stars on prograde
($V_{\phi}$ or $J_{\phi} >$~0) 
and retrograde 
($V_{\phi}$ or $J_{\phi} <$~0) orbits,
and the net rotation for this sample of 35~stars is consistent with zero
(unweighted mean $V_{\phi} = 7 \pm 24$~\kmsec, with 
standard deviation 142~\kmsec).
The relative balance in these quantities 
would indicate that phase-mixing has occurred
among an accreted population.

\begin{figure}
\begin{center}
\includegraphics[angle=0,width=3.35in]{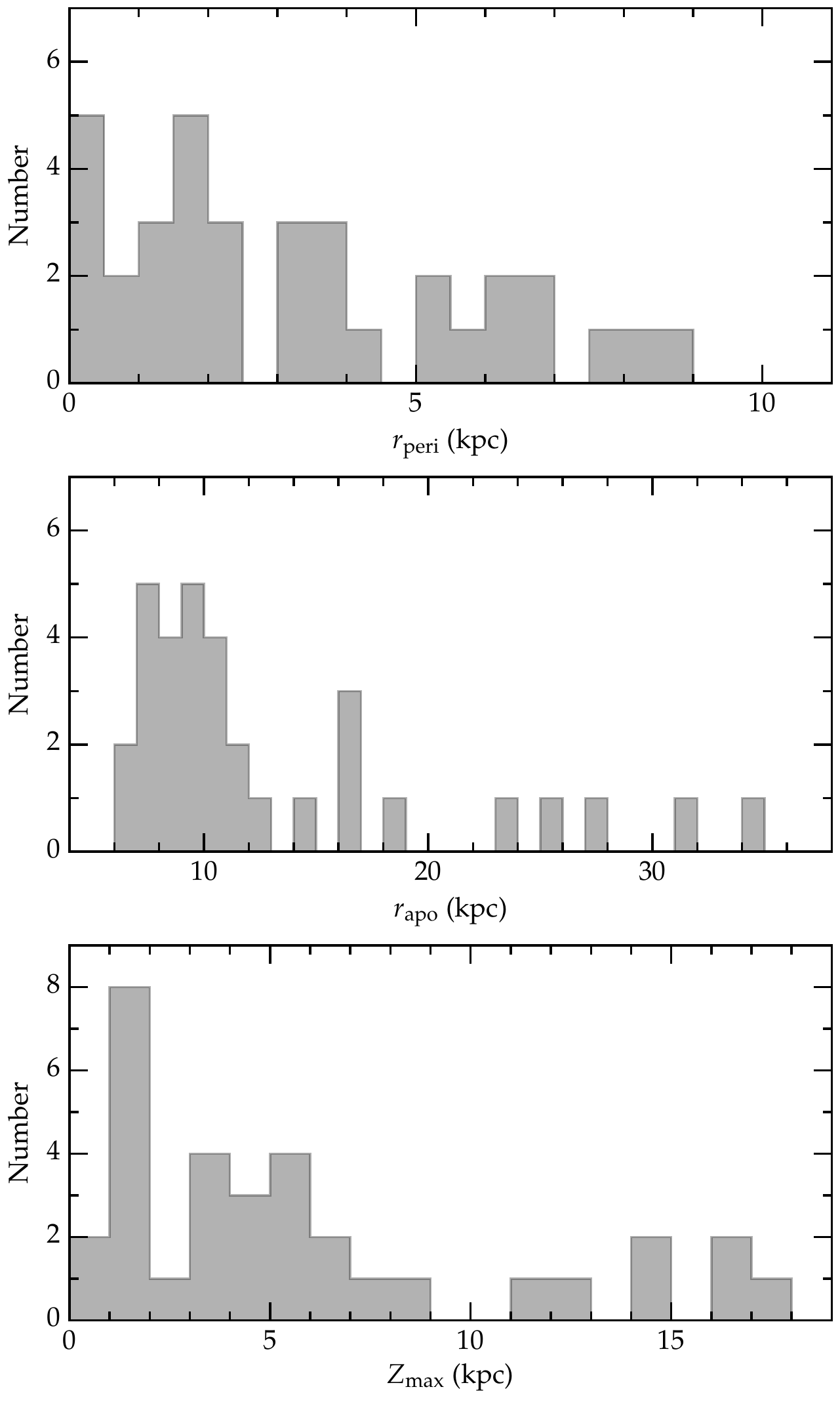}
\end{center}
\caption{
\label{periapoplot}
Histograms of the pericentric radii (top),
apocentric radii (middle), and maximum distance
above or below the Galactic plane (bottom)
for the 35 highly \rpro-enhanced stars in our sample.
Note that two stars with large $r_{\rm apo}$
($63^{+35}_{-15}$~kpc and
$434^{+140}_{-138}$~kpc)
and $Z_{\rm max}$
($48^{+22}_{-10}$~kpc and
$151^{+61}_{-53}$~kpc)
are not shown in the middle and bottom panels.
}
\end{figure}

Most of the stars in our sample
always remain in the inner regions of the 
Galactic halo.
Figure~\ref{periapoplot} shows histograms of their
$r_{\rm peri}$, $r_{\rm apo}$, and $Z_{\rm max}$ values.
Many of the stars (77\%) are on eccentric
($e >$~0.5), radial orbits, and
66\% of the stars have $r_{\rm apo} <$~13~kpc,
indicating that they are at or near apocenter 
while in the Solar Neighborhood.
More than 51\% of the stars pass within 2.6~kpc
of the Galactic center at pericenter, and 20\%
of the sample passes within 1~kpc.
Most of the stars (71\%)
travel at least 3~kpc
above or below the Galactic plane,
and 20\% of the stars orbit beyond 
$r_{\rm apo}$ or $Z_{\rm max} >$~20~kpc.

The star in our sample with the highest energy, 
\object[SMSS J024858.41-684306.4]{SMSS~J024858.41$-$684306.4}
([Fe/H]~$= -$3.71, [Eu/H]~$= -$2.71, [Eu/Fe]~$= +$1.00; 
\citealt{jacobson15smss}),
is only loosely bound to the Milky Way.
Its orbital properties lie well beyond
several of the axes shown in Figure~\ref{fehgroupplot},
with 
$r_{\rm apo} = 434^{+140}_{-138}$~kpc and
$Z_{\rm max} = 151^{+61}_{-53}$~kpc.
This orbit extends well beyond the Milky Way virial radius
($\approx$~280~kpc; \citealt{blandhawthorn16}),
and it may be on its first infall.
Another star,
\object[BPS CS 29491-069]{CS~29491-069}
([Fe/H]~$= -$2.55, [Eu/H]~$= -$1.59, [Eu/Fe]~$= +$0.96; 
\citealt{hayek09}),
also has orbital properties too large for 
Figure~\ref{fehgroupplot}, with
$r_{\rm apo} = 63^{+35}_{-15}$~kpc and
$Z_{\rm max} = 48^{+22}_{-10}$~kpc.
While our work was in the final stages of preparation,
\citet{hawkins18} presented an abundance analysis
of several high-velocity stars, including two stars
with [Eu/Fe]~$\geq +$0.7 that appear to have \rpro\ signatures.
The kinematics of one of these stars, 
\textit{Gaia}~DR2~2233912206910720000
([Fe/H]~$= -$1.72, [Eu/H]~$= -$0.61, [Eu/Fe]~$= +$1.11),
were examined by \citet{hattori18hvs}, who
found that it
has only a 16\% chance of being bound to the Milky Way.
Whether or not it is bound, it has a high eccentricity
and travels several hundred kpc from the Galactic center.
High levels of \rpro\ enhancement 
may be found in stars formed with a wide range of 
initial separations from the Milky Way, not just those
that formed in the inner regions of the halo.

\subsection{Clustering Analysis}
\label{clustering}

In this section
we investigate whether any subsets of 
highly \rpro-enhanced stars could have 
formed together in individual satellites that were subsequently
disrupted by the Milky Way.
Simulations of satellite disruption
indicate that structure remains in phase space
after many Gyr of evolution
\citep{helmi00,font06phase,gomez10}.
This structure may be distinguished in
a Lindblad diagram
($E$ versus $L_{z}$, the $z$ component of the angular momentum), 
despite the fact that 
individual particles would be
smoothly distributed across the sky when viewed from the Solar Neighborhood.
Particles from an individual satellite 
are not single-valued in $E$ or $L_{z}$, but they
exhibit a small, characteristic spread
(e.g., figure~4 of \citeauthor{gomez10}).
\citet{jeanbaptiste17} raised concerns about
using kinematics as the only tracer of 
structure in phase space,
because multiple structures can overlap.
Furthermore, structure may not be uniquely identified
with an accreted component
in galaxies with active merger histories and
relatively massive companions
with small pericentric radii.
The Milky Way's merger history is not so active,
and our study alleviates these concerns by
using a chemically-selected sample.

We search for structure among the 
highly \rpro-enhanced stars in our sample
using energy ($E$) and actions 
($J_{r}$, $J_{\phi}$, $J_{z}$).
Energy is conserved as long as 
the potential of the Milky Way is static,
and \citet{gomez10} showed that 
stars stripped from the same satellite
remain clumped in $E$ even in a realistic time-dependent potential.
The actions are insensitive to the slow, adiabatic
time-dependence of the potential. 
The azimuthal action $J_{\phi}$ ($=-L_{z}$) 
is conserved even if the potential is rapidly changing,
as long as the potential remains axisymmetric. 

We apply four clustering methods 
using the implementation from the 
\texttt{scikit-learn} python package:\
K-means \citep{lloyd82,arthur07},
agglomerative \citep{ward63},
affinity propagation \citep{frey07}, 
and 
mean-shift clustering \citep{comaniciu02}.
All four methods automatically assign each star
to be a member of a cluster, although a cluster
may consist of only one star.
We adjust the clustering parameter values
(number of clusters to be found, metric to be minimized, etc.)
so that the resulting clusters appear reasonably coherent.
Quantitative evaluations of the clusters
(the silhouette coefficient, \citealt{rousseeuw87};
or the Calinski-Harabasz Index, \citealt{calinski74})
fail to distinguish precise values for the number of clusters
or other parameters beyond general ranges.
Typically $\approx$~8--12 clusters are preferred,
including single-star clusters.
We test the reproducibility of these clustering methods 
by repeatedly drawing from the input error distributions
and recomputing the clusters.
The K-means method is highly sensitive to these draws, resulting in
unstable cluster membership from one draw to another, 
so we discard the K-means method from further consideration.
The other three methods are stable against the input draws,
with only 1--2 stars changing cluster membership 
$<$~10\% of the time.

\begin{figure*}
\begin{center}
\includegraphics[angle=0,width=6.0in]{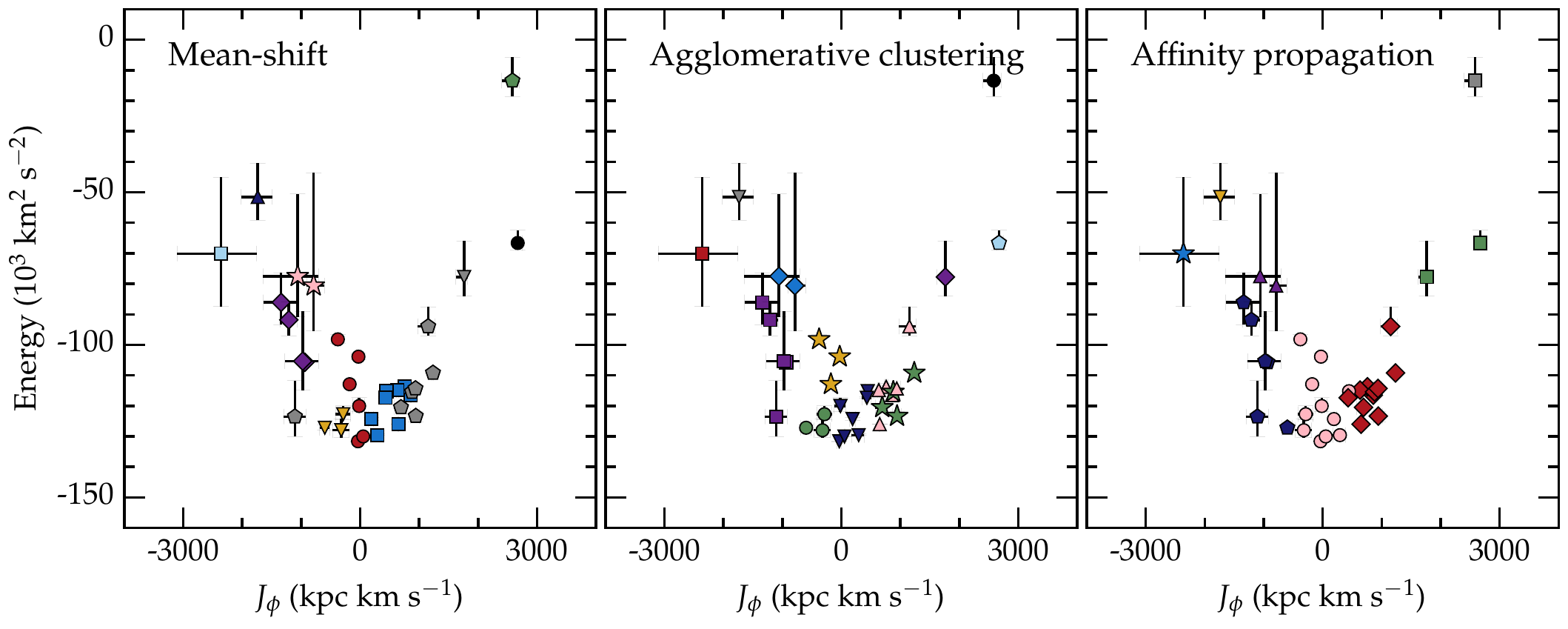} 
\end{center}
\caption{
\label{clusterplot}
Results of three different clustering methods
applied to the 35~stars in our sample.
Different symbol shapes and colors in each panel
indicate the different groups.
Each symbol shape and color is only used in a single panel,
and these combinations are not used in other figures.
The error bars represent the 16$^{\rm th}$ and 84$^{\rm th}$
percentiles of the distributions.
Only the relationship between $E$ and $J_{\phi}$ is shown,
even though the clustering has been performed over
four dimensions ($E$, $J_{r}$, $J_{\phi}$, and $J_{z}$).
}
\end{figure*}

Figure~\ref{clusterplot} illustrates our adopted
clustering results from the three methods.
Only the relationship between $E$ and $J_{\phi}$ is shown,
although the clustering has been performed for 
$E$, $J_{r}$, $J_{\phi}$, and $J_{z}$.
Each set of symbols in each panel represents one cluster.
A few points are worth mentioning.
First,
the stars with lower values of $E$ 
overlap more than stars with higher $E$, so
cluster definitions are more challenging
in this region of the diagram.
Secondly,
there are relatively few
stars with high $E$ values,
so these stars are commonly identified
as the single members of their clusters.
Thirdly,
although the membership of individual clusters 
differs from one method to another,
some subsets of stars are always grouped together.

\begin{figure*}
\begin{center}
\includegraphics[angle=0,width=6.0in]{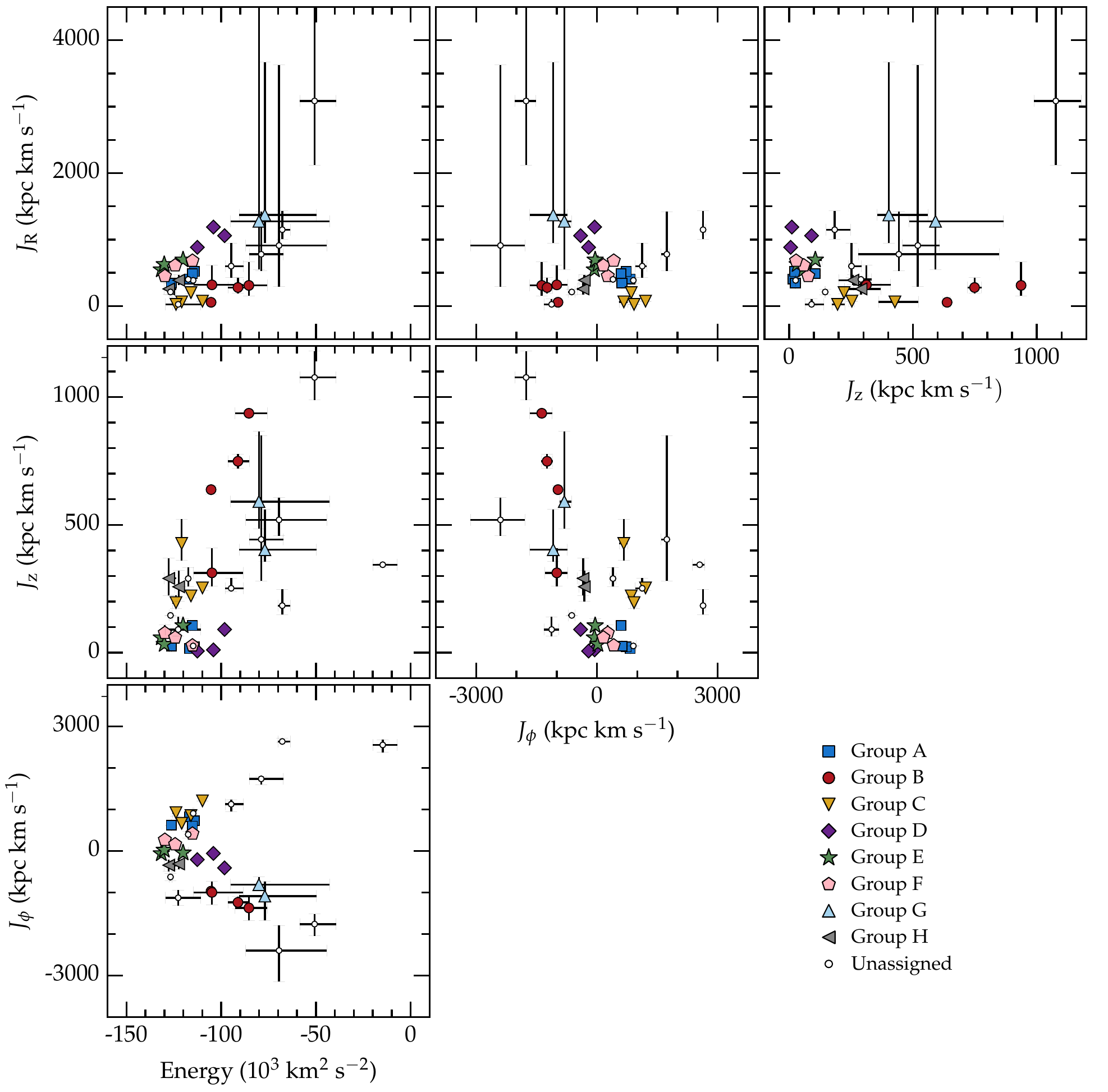} 
\end{center}
\caption{
\label{energyactiongroupplot}
Groups of field \rpro-enhanced stars
as functions of 
$E$, $J_{r}$, $J_{\phi}$, and $J_{z}$.
Each group of stars is indicated by a different symbol/color
combination, as indicated in the legend.
The small white circles indicate stars that 
were not conclusively associated with one of these
groups, either
because their membership was ambiguous
or because they are outliers.
The error bars represent the 16$^{\rm th}$ and 84$^{\rm th}$
percentiles of the distributions for each quantity.
}
\end{figure*}

We associate 25 of the 35~stars into eight groups 
for which the three clustering methods all agree.
These groups range in membership from two to four stars each,
and they are
illustrated in Figures~\ref{fehgroupplot} and 
\ref{energyactiongroupplot}
and listed in Table~\ref{grouptab}.
We assert that
the cluster candidates illustrated in Figure~\ref{energyactiongroupplot}
offer a more reprsentative, conservative expression of the data
than the results of any individual clustering method.
The behavior of the candidate clusters resembles
the extended structures in $E$ and $J_{\phi}$ (or $L_{z}$)
predicted by simulations,
as shown in figure~4 of \citet{gomez10}.
The 10~stars 
that could not be conclusively assigned to these groups
are shown as small white circles in Figure~\ref{energyactiongroupplot}.

\begin{deluxetable}{llccc}
\tablecaption{Groups of \rpro\ Enhanced Field Stars
\label{grouptab}}
\tablewidth{0pt}
\tabletypesize{\scriptsize}
\tablehead{
\colhead{Group} &
\colhead{Members} &
\colhead{[Fe/H]} &
\colhead{[Eu/Fe]} &
\colhead{[Eu/H]} 
}
\startdata
A & \object[HE 0430-4901]{HE~0430$-$4901}                          & $-$2.72 & $+$1.16 & $-$1.56  \\
  & \object[BPS CS 22958-052]{CS~22958-052}                        & $-$2.42 & $+$1.00 & $-$1.42  \\
  & \object[HD 221170]{HD~221170}                                  & $-$2.18 & $+$0.80 & $-$1.38  \\
  & \object[2MASS J15271353-2336177]{J15271353$-$2336177}          & $-$2.15 & $+$0.70 & $-$1.45  \\
\hline                                                             
B & \object[SDSS J235718.91-005247.8]{J235718.91$-$005247.8}       & $-$3.36 & $+$1.92 & $-$1.44  \\
  & \object[BPS CS 31082-001]{CS~31082-001}                        & $-$2.90 & $+$1.63 & $-$1.27  \\
  & \object[SMSS J051008.62-372019.8]{SMSS~J051008.62$-$372019.8}  & $-$3.20 & $+$0.95 & $-$2.25  \\
  & \object[HE 1430+0053]{HE~1430$+$0053}                          & $-$3.03 & $+$0.72 & $-$2.31  \\
\hline                                                             
C & \object[2MASS J21064294-6828266]{J21064294$-$6828266}          & $-$2.76 & $+$1.32 & $-$1.44  \\
  & \object[BPS CS 22945-017]{CS~22945-017}                        & $-$2.73 & $+$1.13 & $-$1.60  \\
  & \object[2MASS J15582962-1224344]{J15582962$-$1224344}          & $-$2.54 & $+$0.89 & $-$1.65  \\
  & \object[BPS CS 22943-132]{CS~22943-132}                        & $-$2.67 & $+$0.86 & $-$1.81  \\
\hline                                                             
D & \object[HD 222925]{HD~222925}                                  & $-$1.47 & $+$1.33 & $-$0.14  \\
  & \object[HD 20]{HD~20}                                          & $-$1.58 & $+$0.80 & $-$0.78  \\
  & \object[2MASS J01530024-3417360]{J01530024$-$3417360}          & $-$1.50 & $+$0.71 & $-$0.79  \\
\hline                                                             
E & \object[BPS CS 22945-058]{CS~22945-058}                        & $-$2.71 & $+$1.13 & $-$1.58  \\
  & \object[BD+17 3248]{BD $+$17\degree3248}                       & $-$2.06 & $+$0.91 & $-$1.15  \\
  & \object[2MASS J00405260-5122491]{J00405260$-$5122491}          & $-$2.11 & $+$0.86 & $-$1.25  \\
\hline                                                             
F & \object[HE 2224+0143]{HE~2224$+$0143}                          & $-$2.58 & $+$1.05 & $-$1.53  \\
  & \object[BPS CS 29529-054]{CS~29529-054}                        & $-$2.75 & $+$0.90 & $-$1.85  \\
  & \object[HD 115444]{HD~115444}                                  & $-$2.96 & $+$0.85 & $-$2.11  \\
\hline                                                             
G & \object[BPS CS 22953-003]{CS~22953-003}                        & $-$2.84 & $+$1.05 & $-$1.79  \\
  & \object[HE 2327-5642]{HE~2327$-$5642}                          & $-$2.78 & $+$0.98 & $-$1.80  \\
\hline                                                             
H & \object[2MASS J21091825-1310062]{J21091825$-$1310062}          & $-$2.40 & $+$1.25 & $-$1.15  \\
  & \object[2MASS J19232518-5833410]{J19232518$-$5833410}          & $-$2.08 & $+$0.76 & $-$1.32  \\
\enddata 
\end{deluxetable}

\subsection{Evaluating the Groups Using Stellar Abundances}
\label{chemicalgroups}

No chemical information is considered
in the clustering process,
so the stellar abundances can be used
to evaluate group membership.
Visual inspection of Figure~\ref{fehgroupplot} suggests that
the groups span a small range of [Fe/H] values
relative to the full sample of 35 stars
($\approx$~2.2~dex).
The group [Fe/H] ranges span 0.10 to 0.65~dex
with dispersions 0.04 to 0.30~dex.
For comparison, the seven highly \rpro-enhanced stars in \rettwo\ 
whose abundances have been studied
span a range in [Fe/H] of 0.93~dex
with a dispersion of 0.35~dex \citep{ji16ret2}.
The typical statistical (systematic) uncertainties of [Fe/H] ratios
are $\approx$~0.1 (0.2) dex,
so the smallest group [Fe/H] dispersions 
may represent upper limits.
The small dispersions
could signal that our groups represent remnants of
individual dwarf galaxies
with moderate spreads in [Fe/H].

\begin{figure}
\begin{center}
\includegraphics[angle=0,width=3.35in]{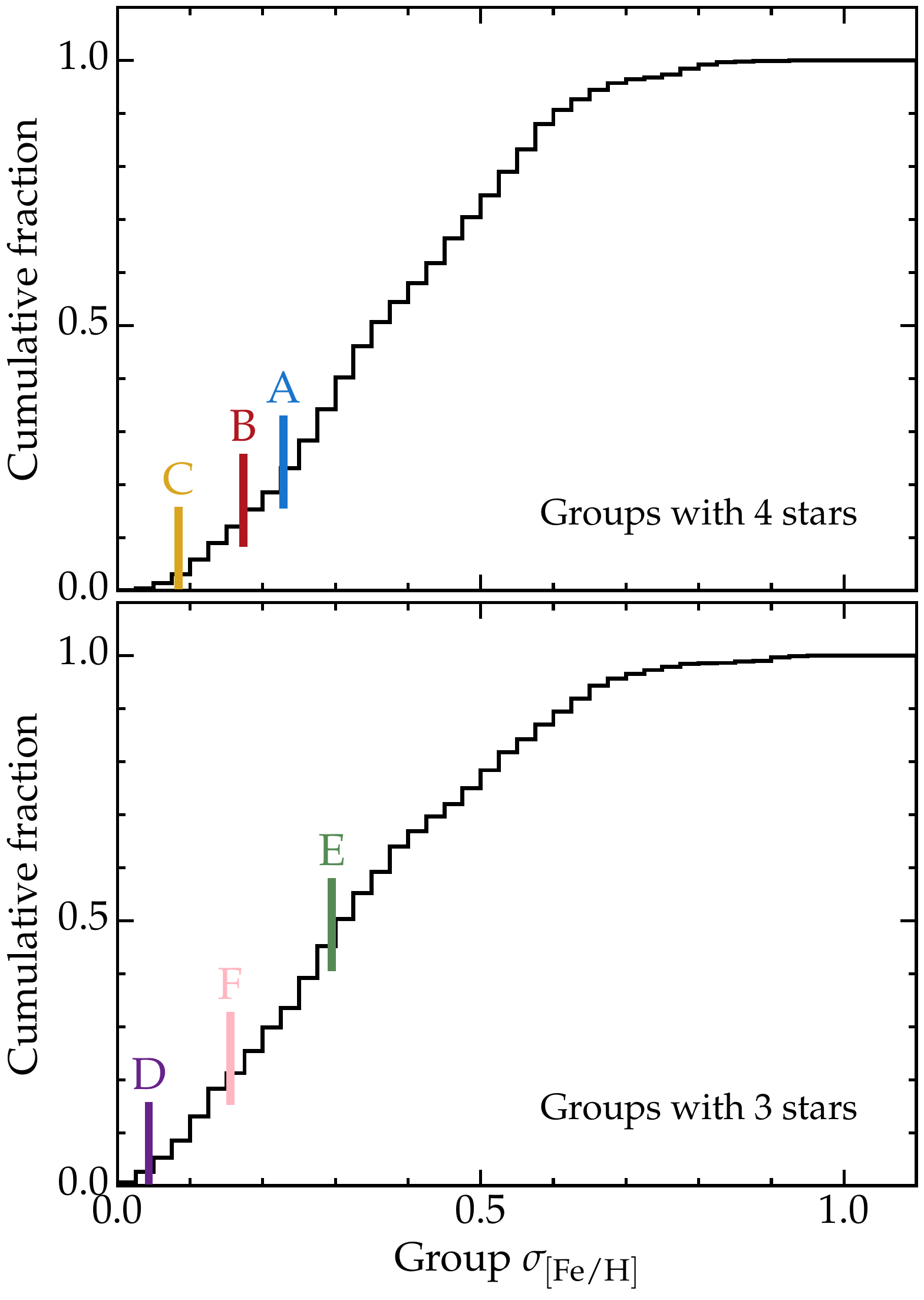}
\end{center}
\caption{
\label{fehsigmaplot}
Comparison of the [Fe/H]
dispersion ($\sigma_{\rm [Fe/H]}$)
for Groups A, B, C, D, E, and F with
cumulative probability density distributions 
for the sample standard deviation of
random draws of [Fe/H] from our sample of 35~stars.
The top panel shows the results of drawing four [Fe/H] values,
and the bottom panel shows the results of drawing three [Fe/H] values.
For comparison, the [Fe/H] dispersion in \rettwo\ is 0.35~dex.
}
\end{figure}

We check the significance of the small [Fe/H] dispersions
for stars within each group as follows.
We draw four [Fe/H] values at random,
without replacement, from the 35~stars 
and compute the sample standard deviation.
We repeat this process 10$^{5}$ times
and compute the probability density distribution.
Figure~\ref{fehsigmaplot} illustrates the results of this test.
The top panel indicates the
[Fe/H] dispersions for the stars in Groups
A, B, and C, each of which contain four stars.
The bottom panel of Figure~\ref{fehsigmaplot} illustrates the
results of an analogous test where three [Fe/H] values are drawn
at random, and these results are compared with
the [Fe/H] dispersions for the stars in Groups D, E, and F,
each of which contain three stars.
All six groups have smaller [Fe/H] dispersions than the median
expected dispersion, and
the dispersions in Groups A--F
are smaller than that of
randomly selected stars in 
80.8\%, 87.9\%, 98.1\%, 98.0\%, 55.6\%, and 80.9\%
of cases, respectively.
The dispersion in [Eu/H] is also small for many of these groups,
and a similar analysis reveals that it is
smaller than that of randomly selected stars in
99.3\%, 29.7\%, 93.5\%, 54.2\%, 74.3\%, and 63.6\% of cases for
Groups A--F, respectively.
We regard this result as evidence that at least some of these
groups are legitimate,
and we proceed under this assumption
as a starting point for investigation.
Additional tests of the legitimacy of these groups,
such as searches for more \rpro-enhanced stars with similar kinematics
or revisiting the clustering analysis using a range of 
Milky Way potentials,
would be most welcome.

\subsection{The Environment of the {\it R}-Process}
\label{environment}

Environmental 
constraints on the \rpro\ have been derived from the
ultra-faint dwarf galaxy \rettwo\ 
(e.g., \citealt{ji16nat,beniamini16b,safarzadeh17ret2})
and chemical evolution models of the Milky Way
(e.g., \citealt{cescutti15,ishimaru15,shen15,cote18rpro}; 
and references therein).
Astrometry from the \textit{Gaia} satellite now enables
similar constraints to be derived from
stars in the Solar Neighborhood.
The highly \rpro-enhanced field stars in our sample
may be in situ halo stars, 
disk or bulge stars ejected into halo orbits, or
accreted from satellites.
A population of stars ejected from the disk would
retain a net prograde rotation.
The net rotation of our sample is consistent with zero
(Section~\ref{orbitresults}), so it is unlikely that
our sample is dominated by stars ejected from the disk.
An accretion origin is likely for the
stars with large orbital apocenters
or retrograde orbits.
In situ or ejected stars are less likely to
exhibit structure in both phase space and metallicity, so
the groups identified in Section~\ref{clustering} also
favor the accretion scenario.
The 25~stars in groups
and the 7 unaffiliated stars with 
$r_{\rm apo} >$~20~kpc or $V_{\phi} <$~0~\kmsec\
constitute the majority (91\%) of the sample,
so most if not all of these highly \rpro\ enhanced stars
were likely accreted.

The metallicity range of our sample 
([Fe/H]~$< -$1.4)
overlaps with the metal-poor end of the Milky Way disk.
A sizable number of stars with [Fe/H]
at least as low as $-$2, and possibly lower,
are found on disk-like orbits 
(e.g., \citealt{ruchti11disk,kordopatis13disk,beers14}).
Many of the known highly \rpro-enhanced stars were identified in 
non-kinematically selected surveys, and
\citet{beers00} and \citet{chiba00} confirm that
these surveys contain
metal-poor disk and halo stars.
If the occurrence frequency of highly \rpro-enhanced stars
is only a function of [Fe/H], then such stars should
also be found among the disk populations.
We find no evidence in our data to support this assertion.
Increasing the number of 
highly \rpro-enhanced stars with [Fe/H]~$> -$2
with well-determined distances and kinematics
would help to strengthen this conclusion.

Many \rpro\ events have occurred 
in the Milky Way disk, bulge, and globular clusters.
This fact is evident from the observation that
stars in these populations
contain substantial amounts of \rpro\ material
([Eu/H]~$> -$1.5), comparable to the
most highly \rpro-enhanced stars in our sample.
The [Eu/Fe] ratios in disk, bulge, and globular cluster stars
are different, however, in that they
rarely exceed [Eu/Fe]~$\approx +$0.6
(e.g., \citealt{gratton04,johnson12,battistini16}).
This observation
suggests that the difference between these populations
and the halo stars we have identified
may be the timescales of producing Fe.
Consequently,
environments with lower star-formation efficiencies
that produce less Fe, like dwarf galaxies, 
may be necessary
to produce extreme ([Eu/Fe]~$\geq +$0.7) \rpro\ enhancement.
This conclusion suggests that the key to obtaining
highly \rpro-enhanced stars may be the environment where
the \rpro\ occurs, not the nature of its site.
A single site, such as neutron-star mergers,
could dominate \rpro\ production in all environments.

\subsection{The Nature of the Progenitor Satellites}
\label{progenitors}

The kinematics and abundances of the accreted stars in our sample 
also help reveal the nature of their progenitor satellites.
Simulations by \citet{wetzel11} and \citet{vandenbosch16}
find that satellites accreted at higher redshifts
are more radial,
have smaller orbital radii, 
and are more tightly bound to the host
than those accreted at lower redshifts.
The simulations of \citet{rocha12} 
revealed that satellites with earliest infall times were
the most tightly bound at $z =$~0.
\citeauthor{rocha12}\ anticipated
that a similar relation for tidal streams could exist,
although dynamical friction can 
complicate this simple picture
(see also \citealt{amorisco17}).
The orbital pericenters of all stars in our sample
are much smaller than that of \rettwo\ 
($29^{+4}_{-6}$~kpc, \citealt{simon18};
 $20 \pm 5$~kpc, \citealt{fritz18})
and most of the surviving low-mass dwarf galaxies,
whose orbital pericenters are located 
$\sim$~10--100~kpc from the Galactic center (\citeauthor{fritz18}).
Many of the highly \rpro-enhanced stars in our sample likely originated 
in progenitor systems with small pericentric radii
that caused them to disrupt much earlier
than the surviving dwarf galaxies.

\begin{figure}
\begin{center}
\includegraphics[angle=0,width=3.35in]{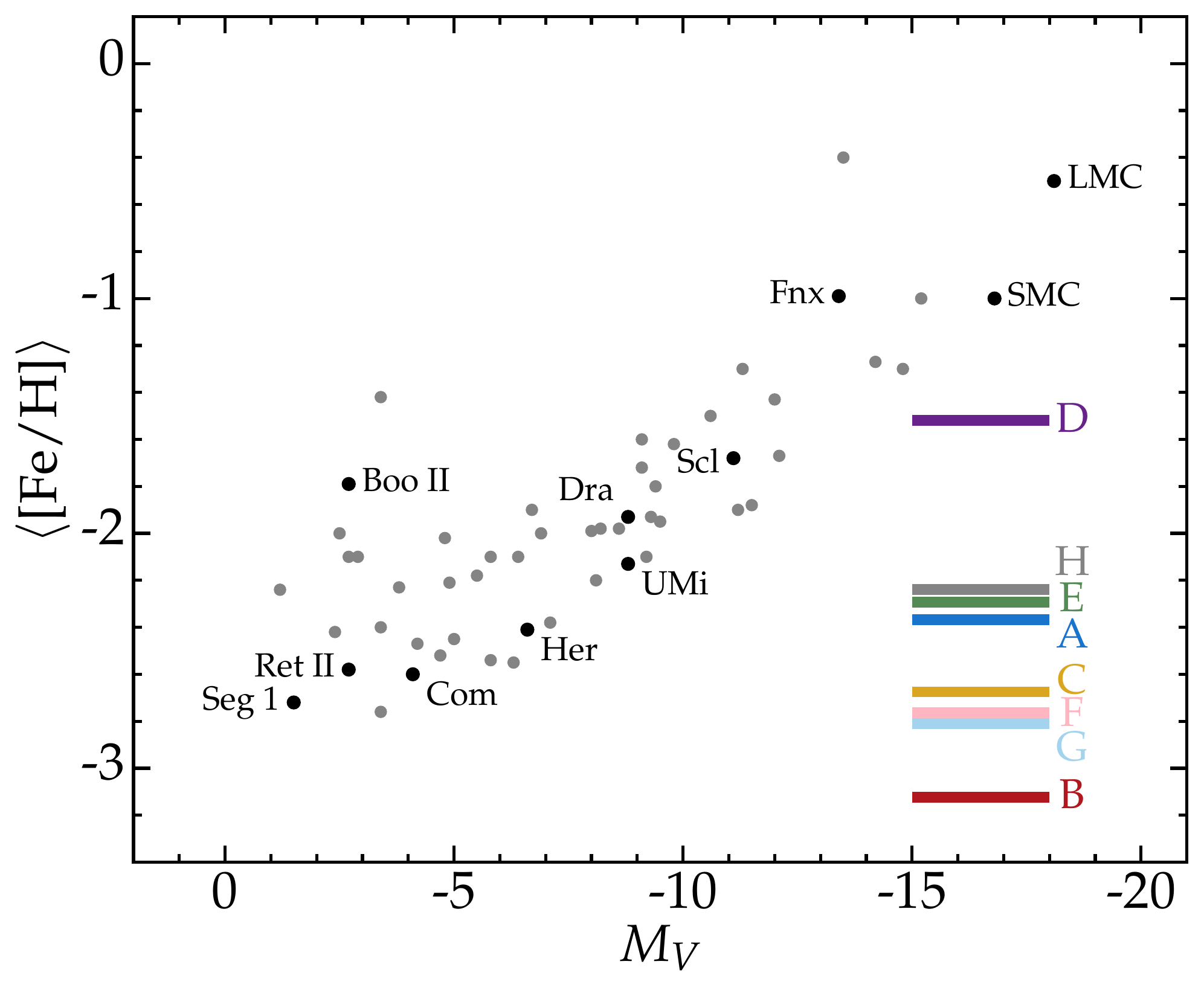}
\end{center}
\caption{
\label{massmetalplot}
Comparison of the mean metallicities of 
Groups A--H with 
the galaxy luminosity-metallicity relation for Local Group galaxies.
Only galaxies with spectroscopically-determined mean metallicities
are shown.
The names of a few representative galaxies are indicated.
The data for the Local Group galaxies are adopted from
\citet{mcconnachie12},
\citet{kirby15hya2,kirby17tri,kirby17leo},
\citet{koposov15dwarfs,koposov15ret2,koposov18hyi},
\citet{laevens15sgr},
\citet{martin15hya,martin16dra},
\citet{walker15ret2,walker16},
\citet{kim16},
\citet{caldwell17},
\citet{carlin17},
\citet{li17eri2},
\citet{simon17},
and
\citet{torrealba18}.
}
\end{figure}

The similarity between the [Eu/Fe] and [Eu/H] ratios found in
\rettwo\ and highly \rpro-enhanced field stars 
(Figure~\ref{euplot})
suggests a common mass scale
between lower-mass dwarf galaxies and 
the progenitor satellites of highly \rpro-enhanced field stars.
Seven of the eight groups we have identified have
mean metallicities $-$3.2~$<$~[Fe/H]~$< -$2.2,
as illustrated in Figure~\ref{massmetalplot}.
If these metallicities are
representative of the progenitor systems, then
the galaxy luminosity-metallicity relation
predicts satellite progenitors with 
$M_{V} \gtrsim -9$ or $\log L \lesssim 5.5$
\citep{kirby08metal,walker16}.
This scale includes ultra-faint dwarf galaxies,
like \rettwo,
and low-luminosity classical dwarf spheroidal galaxies,
like
\object[NAME DRACO DSPH]{Dra} and
\object[NAME URSA MINOR DWARF GALAXY]{UMi},
which are also known to host a handful of 
highly \rpro-enhanced stars
\citep{shetrone01,aoki07th,cohen09dra,cohen10umi,tsujimoto17}.

It seems unlikely that 
the high level of \rpro\ enhancement in Group~D stars
would be found in all stars 
in a progenitor of mass comparable to 
\object[NAME Sculptor Dwarf Galaxy]{Scl}
($M_{*} \approx 2.3 \times 10^{6}$~\msun; \citealt{mcconnachie12}),
because few known field \rpro-enhanced stars 
have such high metallicities.
The small metallicity dispersion of Group~D is
reminiscent of a globular cluster, but
the stars in Group~D do not exhibit the light-element chemical
signatures in O, Na, or Al that are 
found in some globular cluster populations
\citep{barklem05heres,roederer14c,roederer18d}.
We speculate that the stars in Group~D may have formed in a
relatively dense clump of gas in close proximity
to an \rpro\ event in the progenitor system.

We also estimate the initial stellar mass of these systems
using the [Eu/H] ratios and theoretical
\rpro\ yields from neutron-star mergers.
The total stellar mass, $M_{*}$, is given by
\eq{
M_{*} &= 
\varepsilon_{\rm SF} \; 
M_{r, {\rm ej}} \;
f_{r, {\rm Eu}} \; 
A_{\rm Eu}^{-1} \; 
f_{\rm keep} \; 
10^{12-{\rm [Eu/H]}} .
}
We assume that the \rpro\ material is diluted
into H,
but only some fraction of this gas
will be converted into stars.
Here, 
$\varepsilon_{\rm SF}$ is the star formation efficiency,
defined from 0 to 1 as the fraction of gas converted to stars.
This value
is expected to be low in satellites found in low-mass halos
\citep{behroozi13}, and we adopt $\varepsilon_{\rm SF} =$~0.01.
$M_{r, {\rm ej}}$ is the mass in \msun\ of \rpro\ material ejected.
For a neutron-star merger,
$M_{r, {\rm ej}} \sim$~0.005~\msun, which is
the yield of dynamical wind ejecta 
with low electron fractions ($Y_{e}$) that will produce
nuclei at and between the second and third \rpro\ peaks,
including Eu
(see \citealt{cote18rpro}, and references therein).
The term
$f_{r, {\rm Eu}}$ represents 
the mass fraction of Eu among the \rpro\ material,
which is $\approx$~0.006 by mass for nuclei at and beyond the
second \rpro\ peak 
($Z \geq$~52; cf.\ \citealt{roederer12a}) 
when adopting the Solar \rpro\ residuals \citep{sneden08}.
$A_{\rm Eu}$ is the average mass of Eu in atomic mass units,
which is $\approx$~152 for the Eu isotopic ratios
found in the Sun and \rpro\ enhanced metal-poor stars
(e.g., \citealt{sneden02}).
The term
$f_{\rm keep}$ is the fraction of \rpro\ material retained by the system,
which depends on factors like
the location of the event within the 
satellite's potential well \citep{safarzadeh17ret2}.
We adopt $f_{\rm keep} =$~1 
\citep{beniamini18}.
[Eu/H] is the 
average stellar abundance ratio
for each group.

Our calculation makes several simplifying assumptions.
We assume that all stars formed after the injection of \rpro\ material,
which is a reasonable approximation for 
the fraction of stars that are \rpro\ enhanced in \rettwo, 
but it cannot be strictly correct.
We calculate $M_{*}$ values at the time
of star formation, and these values
would be smaller by $\lesssim$~40\% today
because stars with $M \gtrsim$~0.8~\msun\ have evolved and
lost a substantial fraction of their initial mass.
We ignore this small correction when comparing
the initial stellar masses of the satellite progenitors
with the masses of present-day dwarf galaxies.
Under these assumptions, 
the mean [Eu/H] ratios in the eight groups
predict $M_{*} \sim 0.7$--$10 \times 10^{4}$~\msun.
These masses are comparable to the stellar masses of
ultra-faint dwarf galaxies, like
\object[NAME Coma Dwarf Galaxy]{Com}
($M_{*} \approx 7 \times 10^{3}$~\msun) or
\object[NAME Her Dwarf Galaxy]{Her} 
($M_{*} \approx 5 \times 10^{4}$~\msun; \citealt{martin08}).
These order-of-magnitude mass estimates match those from the
luminosity-metallicity relation.

\section{Conclusions}
\label{conclusions}

Traditional chemical tagging relies on the existence 
of chemically homogeneous populations of stars
(e.g., \citealt{freeman02}).
We instead use field stars that are highly enhanced 
in \rpro\ elements ([Eu/Fe]~$\geq +$0.7)
to characterize the environments where the \rpro\ occurred.
We examine the three-dimensional
velocities, integrals of motion, energy, and orbits of
35 highly \rpro-enhanced stars 
with parallax errors $<$12.5\%
in the \textit{Gaia} DR2 catalog.

More than 77\% of the 35 highly \rpro-enhanced stars
are on eccentric ($e >$~0.5), radial orbits.
About 66\% of the stars are on orbits that remain
within the inner regions of the halo ($<$~13~kpc),
and more than 51\% of the stars
pass within 2.6~kpc of the Galactic center.
At the other extreme, 20\% of the stars
have orbital apocenters $>$~20~kpc, including
one star 
(\object[SMSS J024858.41-684306.4]{SMSS~J024858.41$-$684306.4})
whose orbital apocenter is larger than the Milky Way virial radius.
None of the stars have disk-like kinematics,
despite the fact that the metal-rich end of our sample
overlaps with the metal-poor end of the disk.
Roughly equal numbers of stars are moving radially inward and outward,
north and south, and prograde and retrograde,
indicating that a substantial amount of phase-mixing has occurred.

We identify eight candidate
kinematic groups of field \rpro-enhanced stars
based on structure in 
their orbital energies and integrals of motion.
These groups 
show smaller [Fe/H] (and sometimes [Eu/H]) dispersion than 
would be expected by random chance.
The orbital properties, clustering in phase space,
and lack of highly \rpro-enhanced stars on disk-like orbits
indicate that many, if not all,
highly \rpro-enhanced field stars
originated in satellites that were
later disrupted by the Milky Way.
The dwarf galaxy luminosity-metallicity relation
predicts satellite progenitors with 
$M_{V} \gtrsim -9$ or $\log L \lesssim 5.5$
based on the low mean metallicities of seven of the eight groups.
Theoretical \rpro\ yields of neutron-star mergers
and the stellar [Eu/H] ratios 
predict stellar masses 
$\sim 0.6$--$10 \times 10^{4}$~\msun\
for the progenitor systems
when a number of simplifying assumptions are made.
These scales favor ultra-faint dwarf galaxies
or low-luminosity dwarf spheroidal galaxies
as the birthplaces of highly \rpro-enhanced stars.

Comparable levels of \rpro\ enhancement ([Eu/H]~$> -$1.5)
are found in the \rpro-enhanced field stars and
disk and globular cluster populations,
but stars with [Eu/Fe]~$\geq +$0.7 are only
found among stars with halo orbits.
This observation suggests that [Eu/Fe]~$\geq +$0.7 may represent
a more natural lower limit for 
classifying highly \rpro-enhanced stars.
We suggest that the distinguishing factor 
may be the different rates of Fe production.
Environments with lower star-formation efficiency, 
like dwarf galaxies,
may be necessary 
to obtain extreme ([Eu/Fe]~$\geq +$0.7) \rpro\ enhancement
in subsequent stellar generations.
This conclusion allows for the possibility that
a single site, like neutron-star mergers,
could dominate \rpro\ production in all environments.

The significance of our conclusions can be assessed by future work.
Our study is limited by the
small number of highly \rpro-enhanced stars with 
reliable kinematics available at present.
Efforts like the \textit{R}-Process Alliance
will identify much larger samples of such stars in the near future,
and these larger samples will help to confirm or reject the
conclusions of our work.

\acknowledgments

We thank the referee for providing useful suggestions and
members of the stellar halos group at the University of Michigan
for stimulating discussions.
I.U.R.\ thanks T.\ Beers, E.\ Bell, B.\ C\^{o}t\'{e}, G.\ Cescutti, and A.\ Ji
for comments on earlier versions of the manuscript.
K.H.\ thanks E.\ Vasiliev for making Galactic Dynamics code 
{\tt Agama} publicly available. 
I.U.R.\ acknowledges support from
grants AST~16-13536, AST~18-15403,  and
PHY~14-30152 (Physics Frontier Center/JINA-CEE)
awarded by the U.S.\ National Science Foundation (NSF).~
K.H.\ and M.V.\ are
supported by NASA-ATP award NNX15AK79G (PI:\ M.\ Valluri). 
This research has made use of NASA's
Astrophysics Data System Bibliographic Services;
the arXiv pre-print server operated by Cornell University;
the SIMBAD and VizieR
databases hosted by the
Strasbourg Astronomical Data Center;
and
the JINAbase database \citep{abohalima17}.
This work has also made use of data from the European Space Agency (ESA)
mission {\it Gaia} 
(\url{http://www.cosmos.esa.int/gaia}), 
processed by the {\it Gaia} Data Processing and Analysis Consortium 
(DPAC,
\url{http://www.cosmos.esa.int/web/gaia/dpac/consortium}). 
Funding for the DPAC has been provided by national institutions, in particular
the institutions participating in the {\it Gaia} Multilateral Agreement.
Funding for RAVE has been provided by:\ 
the Australian Astronomical Observatory; 
the Leibniz-Institut fuer Astrophysik Potsdam (AIP); 
the Australian National University; the Australian Research Council; 
the French National Research Agency; 
the German Research Foundation (SPP~1177 and SFB~881); 
the European Research Council (ERC-StG 240271 Galactica); 
the Istituto Nazionale di Astrofisica at Padova; 
The Johns Hopkins University; 
the U.S.\ N.S.F.\ (AST-0908326); 
the W.M.\ Keck foundation; 
the Macquarie University; 
the Netherlands Research School for Astronomy; 
the Natural Sciences and Engineering Research Council of Canada; 
the Slovenian Research Agency; 
the Swiss National Science Foundation; 
the Science \& Technology Facilities Council of the UK; 
Opticon; 
Strasbourg Observatory; 
and the Universities of Groningen, Heidelberg and Sydney. 

\facility{%
\textit{Gaia}}

\software{%
Agama \citep{Vasiliev2018},
matplotlib \citep{hunter07},
numpy \citep{vanderwalt11},
scikit-learn \citep{scikitlearn},
scipy \citep{jones01}}

\bibliographystyle{aasjournal}
\bibliography{iuroederer,khattori}

\end{document}